\documentclass[twocolumn,twocolappendix]{aastex631}

\usepackage{soul}

\begin{document}

\title{The Search for Disk Perturbing Planets Around the Asymmetrical Debris Disk System HD 111520 Using REBOUND}

\author[0000-0003-4909-256X]{Katie A. Crotts}
\affiliation{Physics \& Astronomy Department, University of Victoria, 3800 Finnerty Rd. Victoria, BC, V8P 5C2}

\author[0000-0003-3017-9577]{Brenda C. Matthews}
\affiliation{Herzberg Astronomy and Astrophysics, National Research Council of Canada, 5071 West Saanich Rd., Victoria, BC V9E 2E7, Canada}



\begin{abstract}

Debris disks, which are optically thin, dusty disks around main sequence stars, are often found to have structures and/or asymmetries associated with planet-disk interactions. Debris disk morphologies can hence be used as probes for planets in these systems which are unlikely to be detected with other current exoplanet detection methods. In this study we take a look at the very asymmetrical debris disk around HD 111520, which harbours several signs of perturbation such as a ``fork"-like structure in the NW, as well as a 4$^{\circ}$ warp from the midplane on either side of the disk. We simulate the complicated disk morphology using the code REBOUND, with the goal of constraining the possible mass and orbit of the planet responsible for the observed structures. We find that a $\sim$1 M$_{jup}$, eccentric planet that is inclined relative to the disk and has a semi-major axis of $\gtrsim$200 au, is able to reproduce the majority of disk features including the warp, fork and radial extent asymmetry. To create the surface brightness asymmetry, a second eccentric planet is required inside the disk inner edge (50 au), although we are unable to produce the 2 to 1 brightness asymmetry observed, suggesting that a second mechanism may be required. Our work demonstrates how debris disk morphologies alone can be used to learn more about the architecture and evolution of a system as a whole, and can provide planet constraints to determine potential targets for current/future instruments such as JWST/NIRCam and GPI 2.0.   

\end{abstract}

\keywords{circumstellar matter --- scattering}


\section{Introduction} \label{sec:intro}
Debris disks are optically thin, dusty disks that are found around $\sim$20 percent of main sequence stars \citep{Matthews14}. Unlike their predecessors, protoplanetary disks, debris disks are not formed from the primordial material from the initial molecular cloud. Instead, they are formed through the collisions of rocky bodies such as asteroids and comets. To sustain debris disks over millions of years, such collisions must be consistently occurring in order to populate the planetesimal belt with sub-micron to millimeter sized dust grains. This requires that the planetesimals in the disk are stirred so that their orbits are perturbed. Debris disks can be stirred by several mechanisms such as planets, stellar flybys and self-stirring by the planetesimals themselves. Previous studies have shown that most debris disks would have to be unrealistically massive to sufficiently stir the disk (e.g. \citealt{Pearce22, Krivov21}), and stellar flybys are not very common. Planets on the other hand are extremely common and have the ability to sufficiently stir the disk. Therefore, it is possible that the mere existence of a debris disk indicates an underlying exoplanet system. 

Additionally, many of the debris disks that have been spatially resolved in scattered light and thermal emission have shown a variety of morphologies and asymmetric structure such as warps, spirals, eccentric disks, clumps, and brightness asymmetries. Such features are also suspected to be due to planets. For example, \citet{LC16} found that a single 10 M$_{\oplus}$ planet on an eccentric orbit is able to produce multiple different types of morphologies observed in real disks such as the ``Moth" (e.g. HD 61005 and HD 32297; \citealt{Hines07,Schneider14}) and the ``Needle" (e.g. HD 15115; \citealt{Kalas07}). Despite this, only a few spatially resolved disks also have known directly imaged planets, such as HR 8799, $\beta$ Pic and HD 106906 \citep{Marios08,Marois10,Lagrange10,Lagrange19,Bailey14}. In many cases, the known planets have been directly linked to the morphologies of their disk, such as the warp in the $\beta$ Pic disk caused by $\beta$ Pic b \citep{Chauvin12}, and the eccentric disk of HD 106906 caused by HD 106906 b \citep{Nesvold17}. This strengthens the argument that planets are the cause of debris disk structures and asymmetries. We can therefore use their morphologies to estimate the probable masses and orbits of unseen planets in these systems. 

The HD 111520 (HIP 62657) debris disk provides an excellent opportunity to connect unseen planets with disk morphology. The system is located 108.1$\pm$0.2 pc \citep{gaia21} away in the Lower-Centaurus Crux (LCC) group at an approximate age of 15 Myr \citep{Pecaut16}, and has been spatially resolved in scattered light with the Space Telescope Imaging Spectrograph (STIS) on the Hubble Space Telescope (HST) \citep{DP15}, the Gemini Planet Imager (GPI) located on Gemini South \citep{Draper16, Crotts22}, and has also been partially resolved in emission with the Atacama Large Millimeter/Submillimeter Array or ALMA \citep{Lieman16}. While STIS data traces the disk halo, consisting of sub-micron sized dust grains on highly eccentric orbits, GPI data traces the micron-sized dust grains located in the main planetesimal belt (located $\sim$50-110 au from the star). The system has no known planets, however, there are multiple disk structures that suggests their presence. The disk presents one of the largest brightness asymmetries of any debris disk, where the northwest (NW) side is 2 times brighter than the southeast (SE) side seen with GPI, and 5 times brighter seen with STIS \citep{Draper16,Crotts22}. The disk also harbours a radial extent asymmetry where both STIS and GPI observations show the NW side of the disk to be more radially extended than the SE side. Tracing the disk spine of the halo additionally reveals a $\sim$4$^{\circ}$ warp from the midplane on either side of the disk, alongside a bifurcation or ``fork"-like structure in the NW where the disk midplane appears to split into two \citep{Crotts22}. While the top fork component aligns with the 4$^{\circ}$ warp, the bottom fork component aligns with the micron sized grains observed with GPI. 

These structures and asymmetries strongly suggest the presence of an unseen planet. For example, the warp and fork is reminiscent of the warp and second disk component seen in $\beta$ Pic, which again can be directly connected to the planet, $\beta$ Pic b, which is inclined relative to the disk \citep{Chauvin12}. The radial extent and surface brightness asymmetry can also be explained by an eccentric disk induced by a planet on an eccentric orbit, such as in the case of the HD 106906 disk \citep{Nesvold17}. Additionally, other explanations have not been able to replicate the overall HD 111520 disk morphology. A recent massive collision between two rocky bodies is another way to create a fork, warp, eccentric disk and brightness asymmetry as found in \citet{Jones23}. However, in order to create the warp and radial extent asymmetry in the right direction, the collision would need to take place on the SE side of the disk, creating a brightness asymmetry opposite of what is observed, such that the SE side is brighter rather than the NW. This begs the question whether or not planets may be a better explanation for the overall disk morphology.

We attempt to answer this question here by using the n-body simulation code REBOUND \citep{Rein12}. N-body simulation codes provide a useful tool for modelling asymmetric disks, as well as providing constraints on the responsible planet. For example, in the case of the HD 106906 disk, using REBOUND, \citet{Nesvold17} was able to provide constraints on the orbit of HD 106906 b, where they found that the planet would need to be on an eccentric and inclined orbit with $i < 10^{\circ}$. Later constraints placed on the planet's orbit based on observations were found to be consistent with the results from these n-body simulations \citep{Nguyen21}. In this paper, we take a similar approach, where we use REBOUND to simulate the complex morphology of the HD 111520 disk via planet-disk interactions. Our goal is to provide information on whether or not a planet(s) may be responsible for the disk structures, and if so, also provide constraints on the mass and orbit of the planet. These constraints will be useful for determining the feasibility of detection with current and future instruments such as NIRCam on the James Webb Space Telescope (JWST) and GPI 2.0. Our model and simulation are described in Section \ref{sec:model} and our results are discussed in Section \ref{sec:res}. We then further explore potential planet properties in Section \ref{sec:constraints} and discuss the implications of our results in Section \ref{sec:discussion}.

\begin{figure}
    \centering
    \caption{\label{fig:initial_config} Initial orbital configurations and positions of the planets and disk for a planet with $a=250$ au (\textbf{left}) and $a = 40$ au (\textbf{right}). The initial setup is viewed both edge on (\textbf{top}) and face on (\textbf{bottom}).}
    \includegraphics[width=0.47\textwidth]{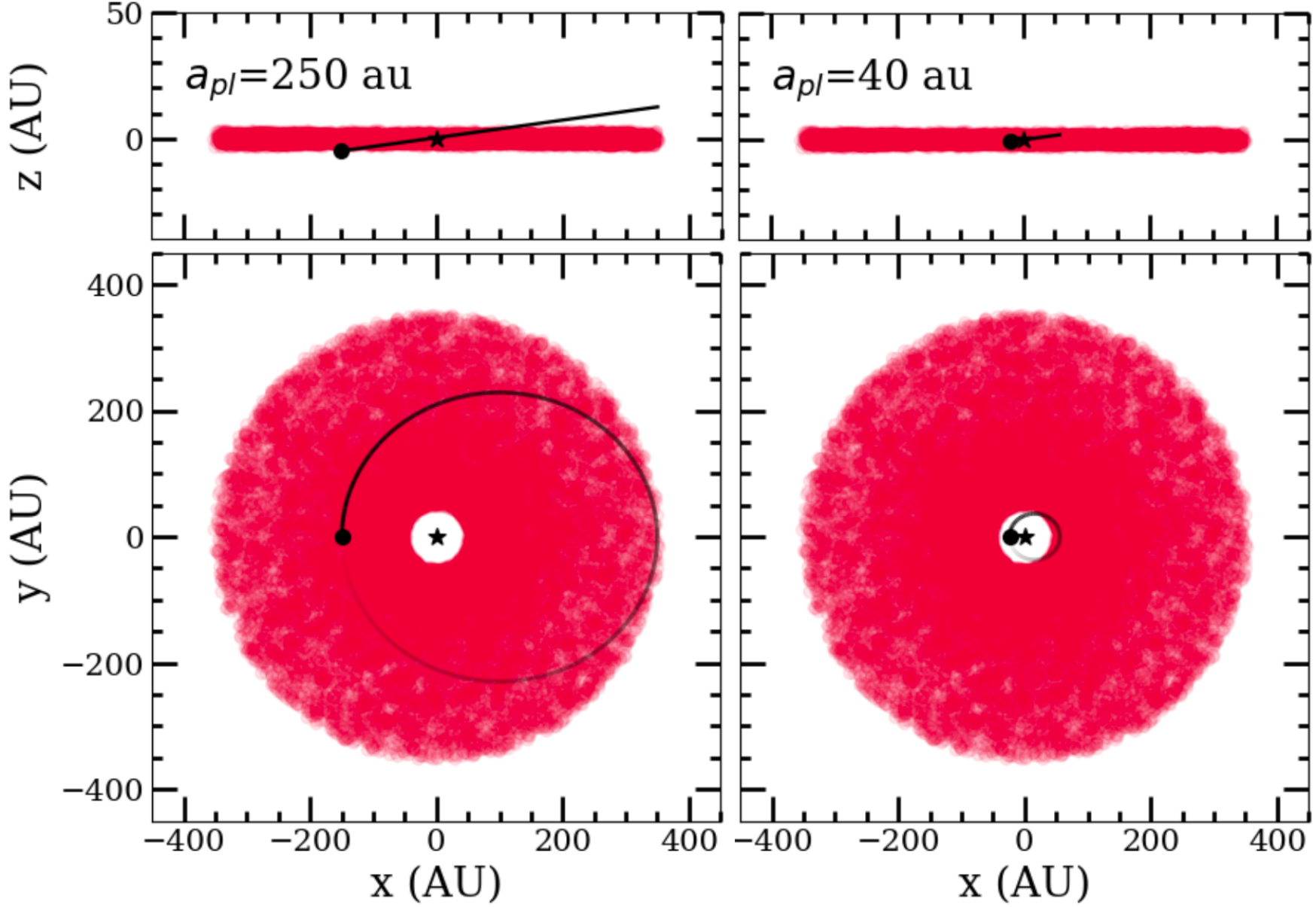}
\end{figure}

\section{Model} \label{sec:model}
We use the n-body simulation code, REBOUND, to create models of the disk with the goal of constraining the planet mass and orbit responsible for the observed morphology. We start by looking at two scenarios: 1) A 1 M$_{jup}$ planet orbiting outside the warp location ($a_{pl}$ = 250), and 2) A 1 M$_{jup}$ planet orbiting inside the disk inner edge ($a_{pl}$ = 40 au). We choose these two scenarios as a planet outside (e.g. HD 106906) and inside the disk inner edge (e.g. $\beta$ Pic) can both perturb the disk through secular perturbations. 250 and 40 au are chosen specifically, as 250 au is near the warp location ($\sim$180 au, \citealt{Crotts22}) and 40 au is near the estimated disk inner edge ($\sim$50 au, \citealt{Draper16}). We also start with a 1 M$_{jup}$ planet as this is below the current upper mass constraint of 3 M$_{jup}$ set by GPI \citep{Nielsen19}, but may be within reach of detection by select instruments such as JWST/NIRCam and GPI 2.0. In both scenarios the planet is on a moderately eccentric ($e$ = 0.4) and inclined ($i = 2^{\circ}$) orbit relative to the disk. The reasoning for these choices is to attempt to replicate the disk halo's significant radial extent asymmetry and the 4$^{\circ}$ warp from the midplane that is observed, as secular perturbations from an eccentric and inclined planet can force planetesimals onto similarly eccentric and inclined orbits (e.g. \citealt{mouillet97,Heap00,Wyatt08}). This results in a disk that is offset from the star (creating a radial extent asymmetry), and can create a warp near the planet location. Additionally, studies have shown that the resulting inclination of the disk $\approx$ 2$i_{pl}$ \citep{Dawson2011}, hence why we choose $i_{pl} = 2^{\circ}$.

\subsection{Simulation Setup}
We start both simulations by adding a 1 solar-mass star, a Jupiter-mass planet, and disk particles. The planet starts at its pericenter, which we set to be on the left side of the star, i.e the argument of pericenter, $\omega_{pl}$, is defined to be $270^{\circ}$. We define $\omega$ in the same way as REBOUND, where $\omega$=0 is in the direction of the observer.

We add 20,000 massless disk particles between 50 and 350 au from the star with random longitude of ascending node ($\Omega_{p}$), longitude of pericenter, and true anomaly ($f_{p}$) between 0 and 2$\pi$. We consider these particles ``parent" particles. We start with a dynamically cold disk where disk particles have a small inclination dispersion of $\pm$0.01 radians (0.58$^{\circ}$), as well as a small eccentricity dispersion of $\pm$0.01. Each particle is also assigned a $\beta$ value, where $\beta$ is the ratio between the force of radiation pressure from the star \citep{Burns79} and the force of gravity ($\beta = F_{rad}/F_{grav}$). Radiation pressure is an important force for debris disks, as dust particles with a $\beta$ greater than 0.5 will be blown out of the system. The $\beta$ value is randomly selected from a distribution between 0.001 and 0.4 with a power law of 3/2 (equivalent to a dust grain size distribution power law of -7/2, \citealt{Dohnanyi69}), i.e. $dN/d\beta \propto \beta^{3/2}$, meaning that the size distribution is dominated by the smallest particles. Using equation 18 from \citet{Wyatt99} and assuming compact astrosilicate grains with a dust particle density of 3.3 g/cm$^{3}$ \citep{Draine03}, our chosen distribution of $\beta$ values correspond to a particle size of $\sim$3 $\mu$m to 970 $\mu$m (0.97 millimeters). For reference, using equation 7 in \citet{pawellek14} yields a blowout size of $\sim$0.75 $\mu$m for the system. The initial configuration of our models can be seen in Figure \ref{fig:initial_config}.

Once all particles are added, the simulation is then integrated over 15 Myr, the estimated age of the system \citep{Pecaut16}, using the Wisdom-Holman integrator, WHFast. Using REBOUNDx \citep{Tamayo20}, radiation pressure is also turned on during this integration period. 

\subsection{Synthetic Scattered Light Images}
To create synthetic scattered light images of our simulations, we follow a similar procedure to other disk simulation papers such as \citet{Nesvold17} and \citet{Moore23}.

Once the simulation is integrated over 15 Myr, we populate the disk with more particles. Since we do not track collisions in the disk, we instead randomly select 500 bound parent particles in the densest regions of the disk (i.e. $e_{p} < 1$ and close to the midplane where collisions are more likely to occur) with a range of true anomalies between 0 and 2$\pi$. We then generate 20 orbits for each parent particle based on their $\beta$ value and orbital properties. Again, $\beta$ for the new dust particles is assigned from a distribution with a power law of 3/2. In this case, the max $\beta$ allowed is determined by the orbit of the parent particle as shown by the following equation:

\begin{equation}
    \beta_{max} = \frac{1 - e_{p}}{2(1 + e_p\text{cos}f_{p})}
\end{equation}

Here, $e_{p}$ is the eccentricity of the parent particle, while $f_{p}$ is the true anomaly of the parent particle. Once $\beta$ is selected for each orbit, the rest of the orbital parameters are calculated including the semi-major axis ($a$), the eccentricity ($e$) and argument of pericenter ($\omega$). Similarly, these parameters are calculated based on the parent orbit using the following equations \citep{Burns79,Wyatt99}:

\begin{equation}
    a = \frac{a_{p}(1 - e_{p}^{2})(1 - \beta)}{1 - e_p^{2} - 2\beta(1 + e_{p}\text{cos}f_{p})}
\end{equation}

\begin{equation}
    e = \frac{\sqrt{e_{p}^{2} + 2\beta e_{p}\text{cos}f_{p} + \beta^{2}}}{1 - \beta}
\end{equation}

\begin{equation}
    \omega = \omega_{p} + \text{tan}^{-1}(\frac{\beta \text{sin}f_{p}}{e_{p} + \beta \text{cos}f_{p}})
\end{equation}

\begin{figure}
    \centering
    \caption{\label{fig:density_1pl} \textbf{Top:} Density map of the 250pl model face on \textbf{Bottom:} Density map of the 40pl model face on. The red points mark the final position of the planet for both models.}
    \includegraphics[width=0.47\textwidth]{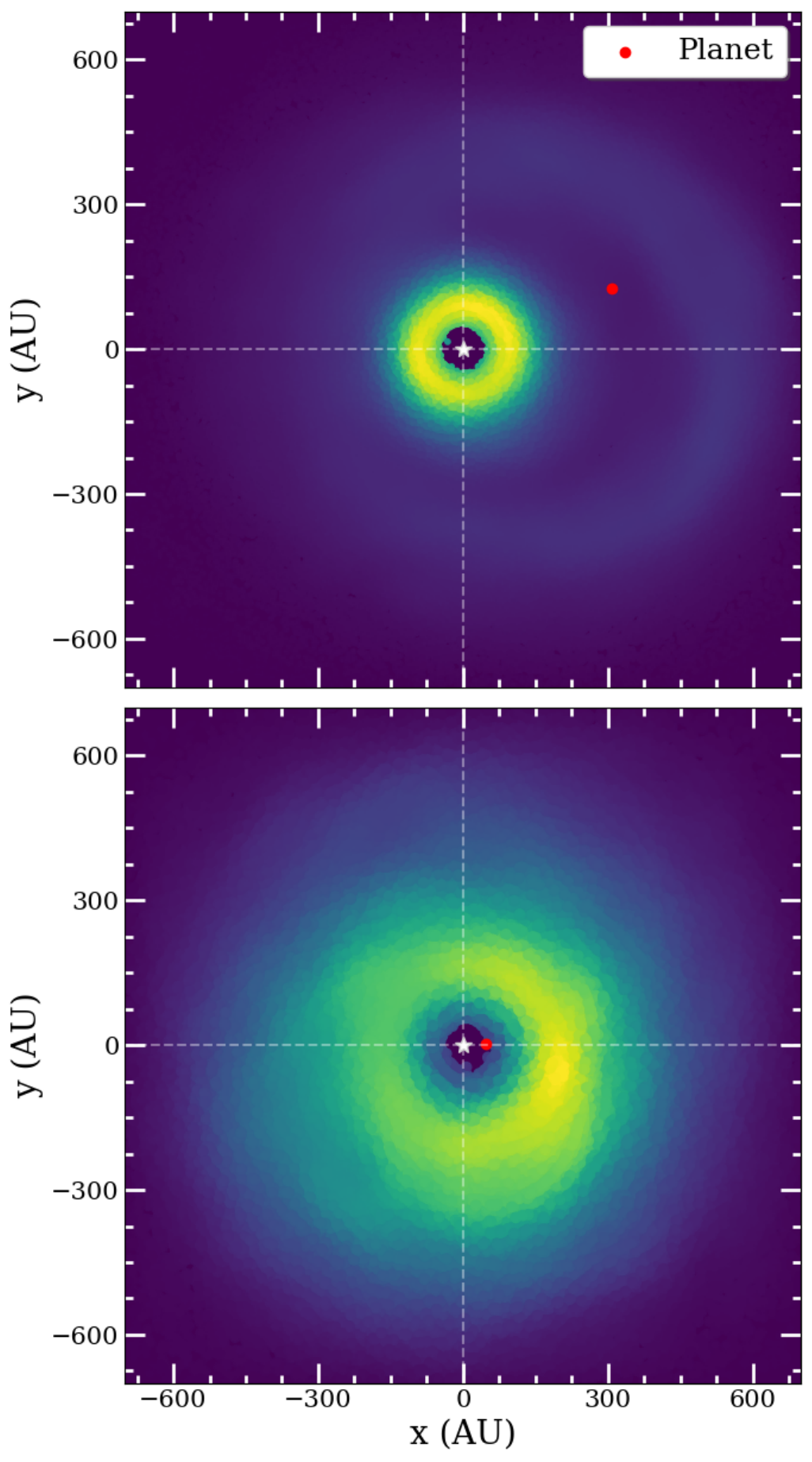}
\end{figure}

Here $a_{p}$ and $\omega_{p}$ are the semi-major axis and argument of pericenter of the parent particle. Besides these parameters, we set the new particle inclination and longitude of ascending node to that of the parent particle ($i$ = $i_{p}$ and $\Omega = \Omega_{p}$). We then generate 20 disk particles per orbit with randomly selected mean anomalies between 0 and 2$\pi$ for a total of 400 disk particles per parent particle. We plot the final density of the two models shown face-on in Figure \ref{fig:density_1pl}.

Once the disk particles are generated, we then apply a Henyey-Greenstein (HG; \citealt{HG41}) SPF to measure the surface brightness of the disk. We first calculate the scattering angle which is defined as the angle between the incident light ray from the star and the scattered light ray from the disk particle in the direction of the observer. The scattering angle is calculated as $\text{cos}^{-1}(z/d)$, where $z$ is the position of the particle in the $z$ direction and $d$ is the total distance of the particle from the star. The scattering angle is then passed through the HG function which is defined as

\begin{equation}
    \phi(g,\theta) = \frac{1}{4\pi} \frac{1 - g^{2}}{(1 + g^{2} - 2g\text{cos}\theta)^{3/2}}
\end{equation}

Where $\theta$ is the scattering angle and $g$ is the asymmetry parameter. We use a 2-parameter HG function ($g_{1}$ and $g_{2}$) as it more accurately represents the SPF of Saturn's rings compared to a single HG function. This SPF has also been found to be similar to the SPF of many debris disks including HD 111520 (\citealt{Hughes18,Hom24}). Based on the SPF measurements done by \citet{HS15} of Saturn's D68 ring, we choose $g_{1}$ = 0.995 and $g_{2}$ = 0.325 with weights $w_{1}$ = 0.779 and $w_{2}$ = 0.221, respectively. The final HG function is the sum of two weighted HG functions calculated with $g_{1}$ and $g_{2}$ (i.e. $\phi(g,\theta) = w_{1}\phi(g_{1},\theta) + w_{2}\phi(g_{2},\theta)$). The final surface brightness is measured as $\phi(g,\theta)/\beta^{2} d^{2}$, where 1/$\beta^{2}$ accounts for the geometric cross section for each disk particle. For our models, we do not consider multiple scattering of photons with the safe assumption that debris disks are optically thin at all wavelengths.

\begin{figure*}[ht!]
    \centering
    \caption{\label{fig:1_planet_mods} \textbf{Top:} Composite image of the HD 111520 disk optical HST/STIS data on large scales and the GPI $H$-band data within the HST coronagraph mask. This figure is modified from \citet{Crotts22}. \textbf{Middle:} Synthetic scattered light image of the HD 111520 disk with 1 planet orbiting outside the warp ($a_{pl}=250$ au) \textbf{Bottom:} Synthetic scattered light image of the HD 111520 disk with 1 planet orbiting inside the disk inner edge ($a_{pl}=40$ au). For both models, the surface brightness is the same in log scale, and surface brightness units are arbitrary. Additionally, the disk is inclined by 89$^{\circ}$. The white dashed lines trace the ``fork"-like structure and warp induced by the planet, and is the same in all three figures.}
    \includegraphics[width=\textwidth]{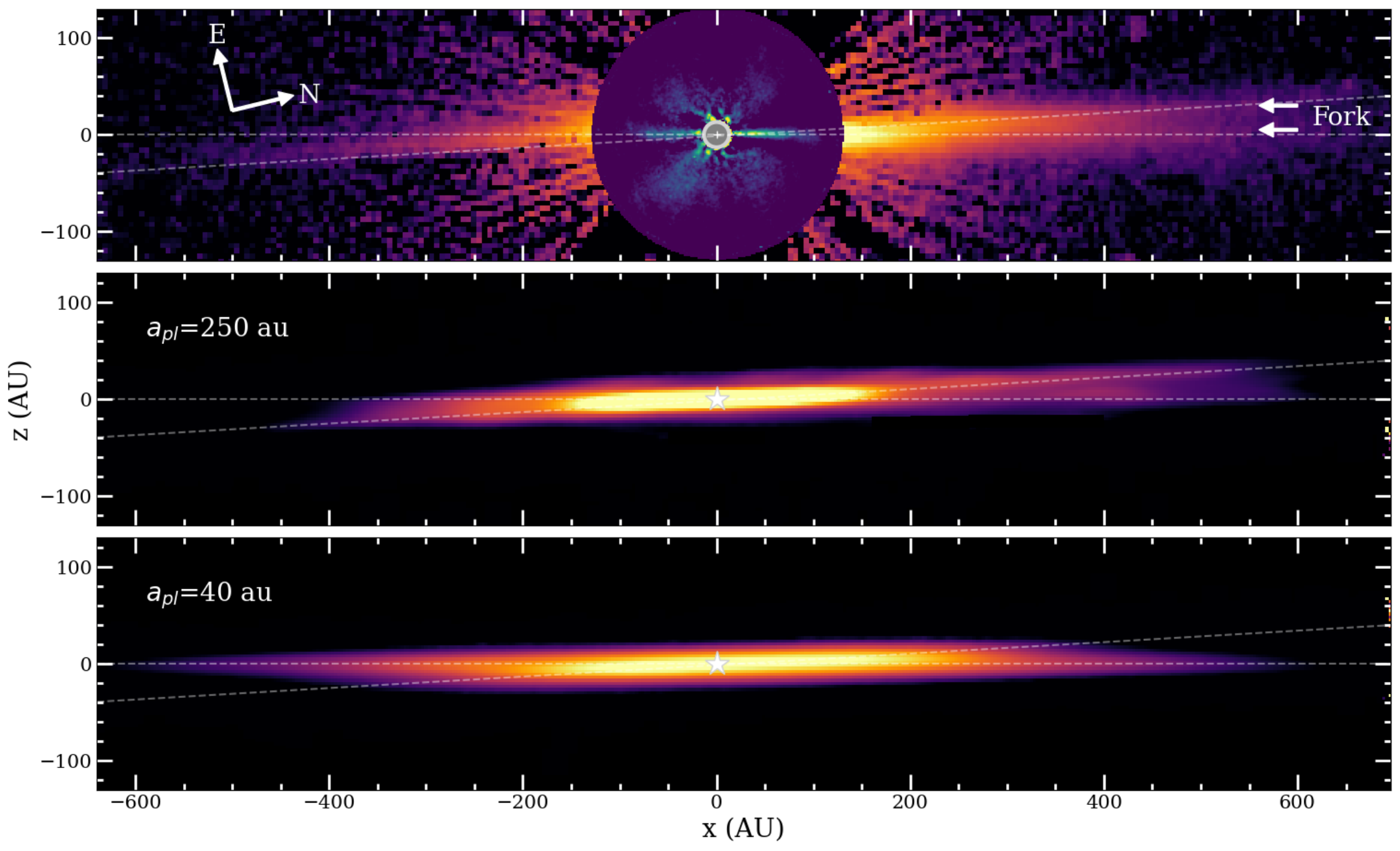}
\end{figure*}

Keeping $\omega_{pl}$ at 270$^{\circ}$, the model is rotated about the x-axis to have an inclination of 89$^{\circ}$ based on empirical measurements of the disk \citep{Crotts22}. The disk particles are then projected onto a 1400 au by 1400 au grid, which we bin into 450, 4 au by 4 au, bins. As a final step, we smooth the image with a Gaussian kernel with $\sigma = 1$ pixel. The final scattered light models for both simulations can be seen in Figure \ref{fig:1_planet_mods}.

\section{Results} \label{sec:res}

We have created two synthetic scattered light models based on REBOUND simulations of a disk with a 1 Jupiter-mass planet with a semi-major axis of 250 au and 40 au, placing the planet slightly outside the warp and inside the disk inner edge. Both planets are on eccentric, inclined orbits relative to the disk. For simplification, from here on forward, we refer to the two models as the 250pl and 40pl models.

\begin{figure*}
    \centering
    \caption{\label{fig:250_rot} Synthetic scattered light image of the HD 111520 disk with 1 planet orbiting outside the warp ($a_{pl}=250$ au). The disk is rotated counter-clockwise by 45$^{\circ}$ intervals starting from $\omega_{d} = 270^{\circ}$ all the way to $\omega_{d} = 225^{\circ}$ \textbf{Left:} Representation of the disk as observed in the optical with HST/STIS \textbf{Right:} Representation of the disk as observed in the NIR with GPI. The rectangle in the top left figure represents the field of view of the NIR models.}
    \includegraphics[width=0.49\textwidth]{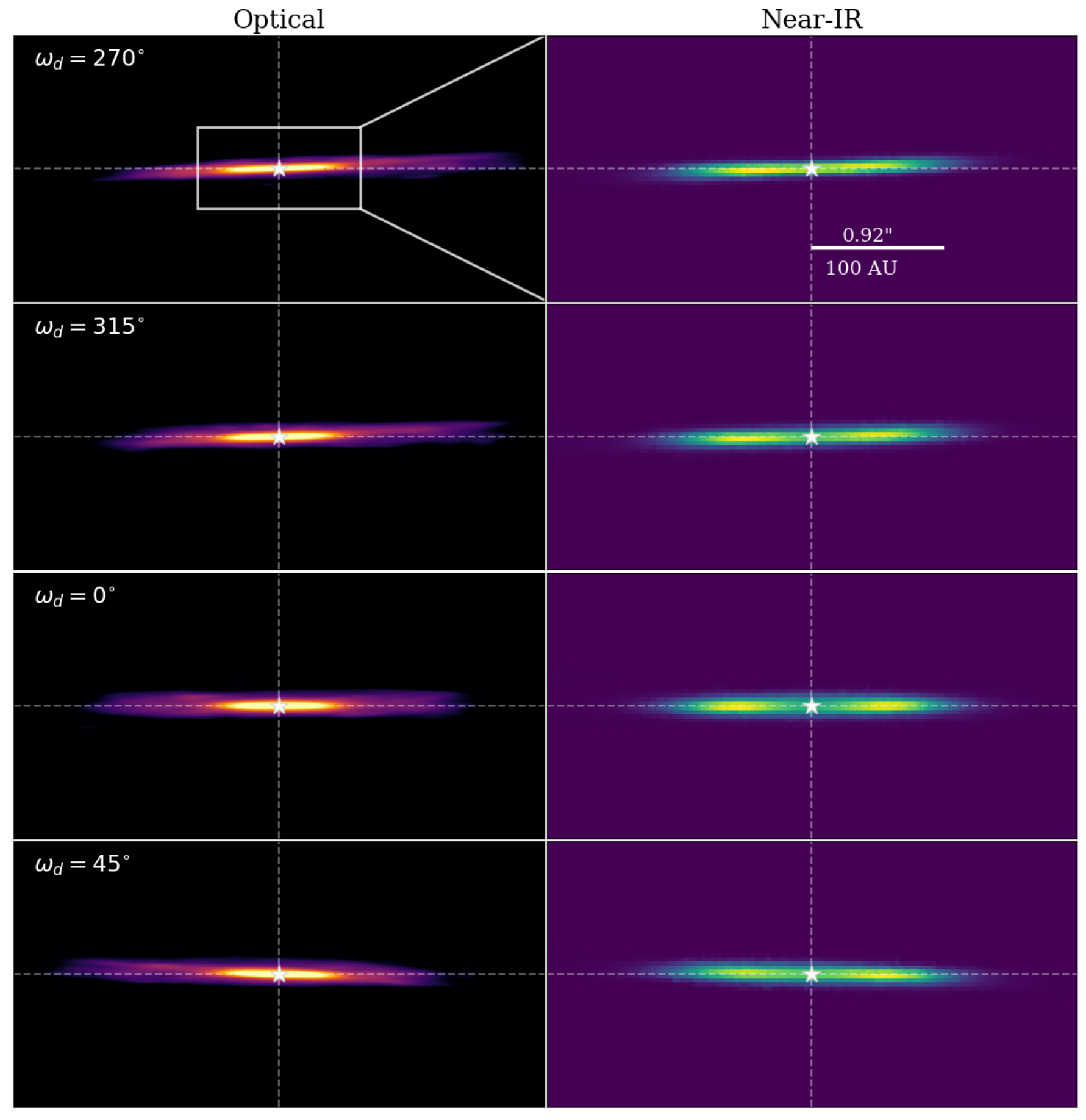}
    \includegraphics[width=0.49\textwidth]{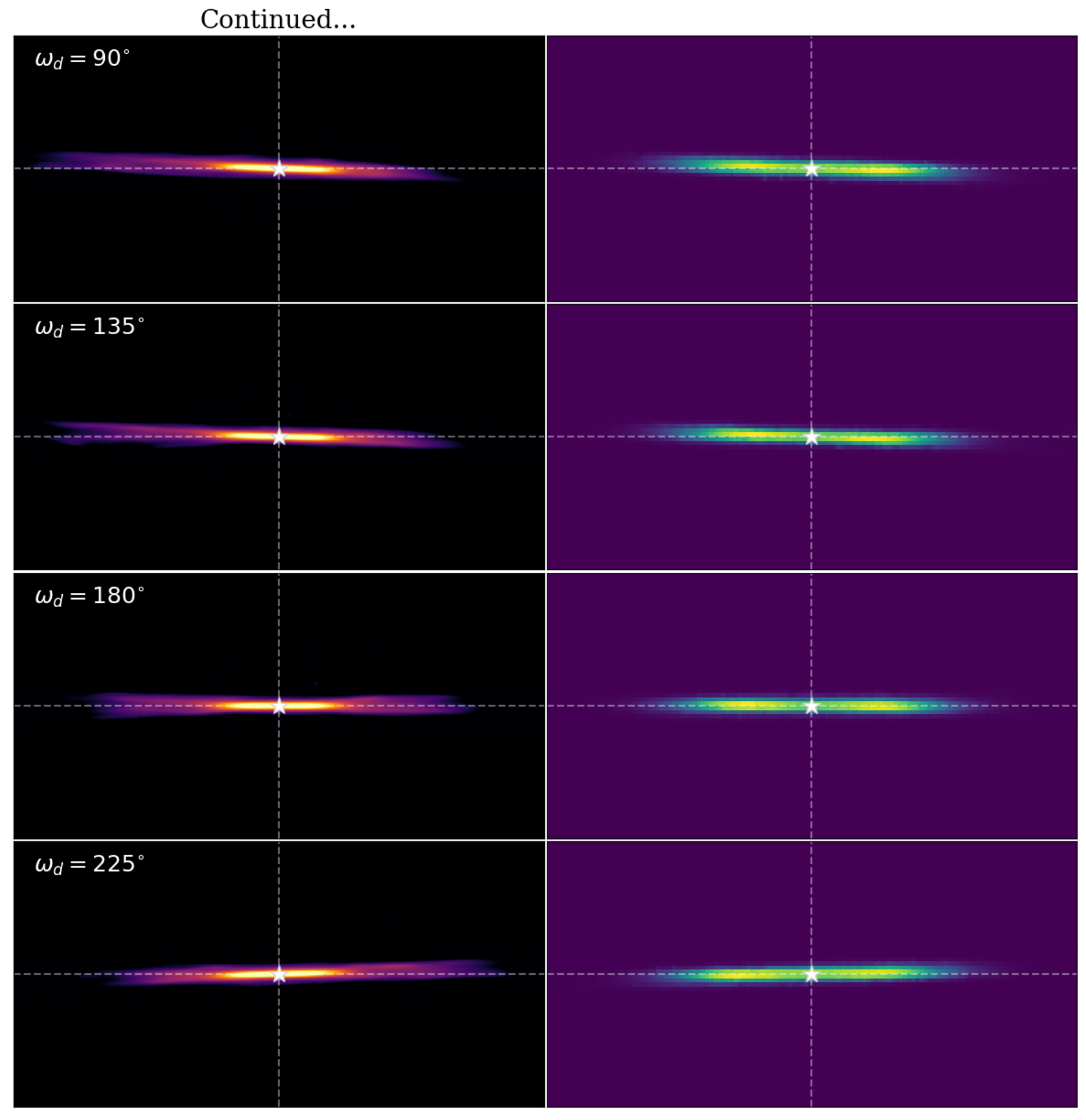}
\end{figure*}

\begin{figure*}
    \centering
    \caption{\label{fig:40_rot} Synthetic scattered light image of the HD 111520 disk with 1 planet orbiting inside the disk inner edge ($a_{pl}=40$ au). The disk is rotated counter-clockwise by 45$^{\circ}$ intervals from $\omega_{d} = 270^{\circ}$ to $\omega_{d} = 225^{\circ}$ \textbf{Left:} Representation of the disk as observed in the optical with HST/STIS \textbf{Right:} Representation of the disk as observed in the NIR with GPI. The rectangle in the top left figure represents the field of view of the NIR models.}
    \includegraphics[width=0.49\textwidth]{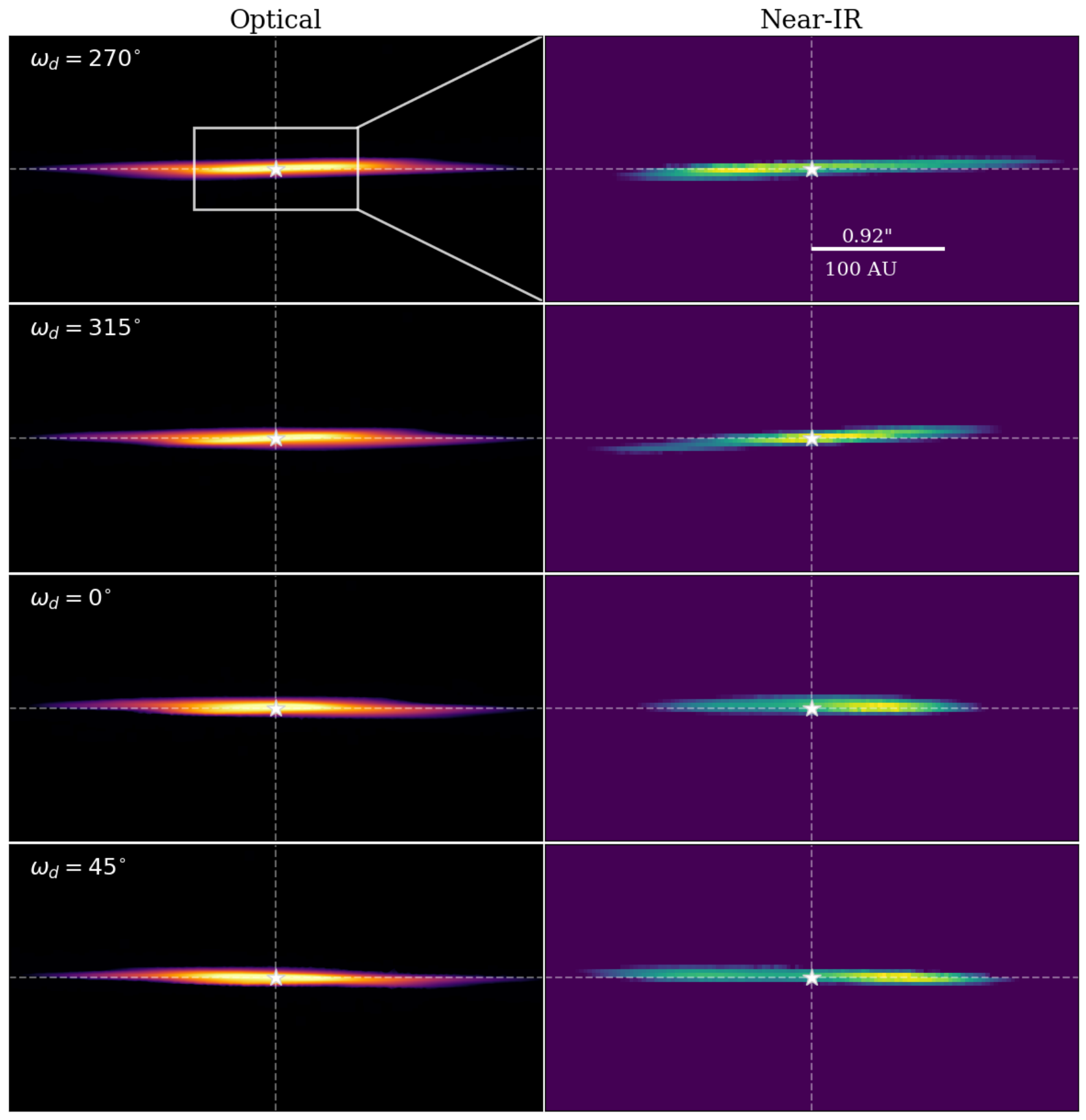}
    \includegraphics[width=0.49\textwidth]{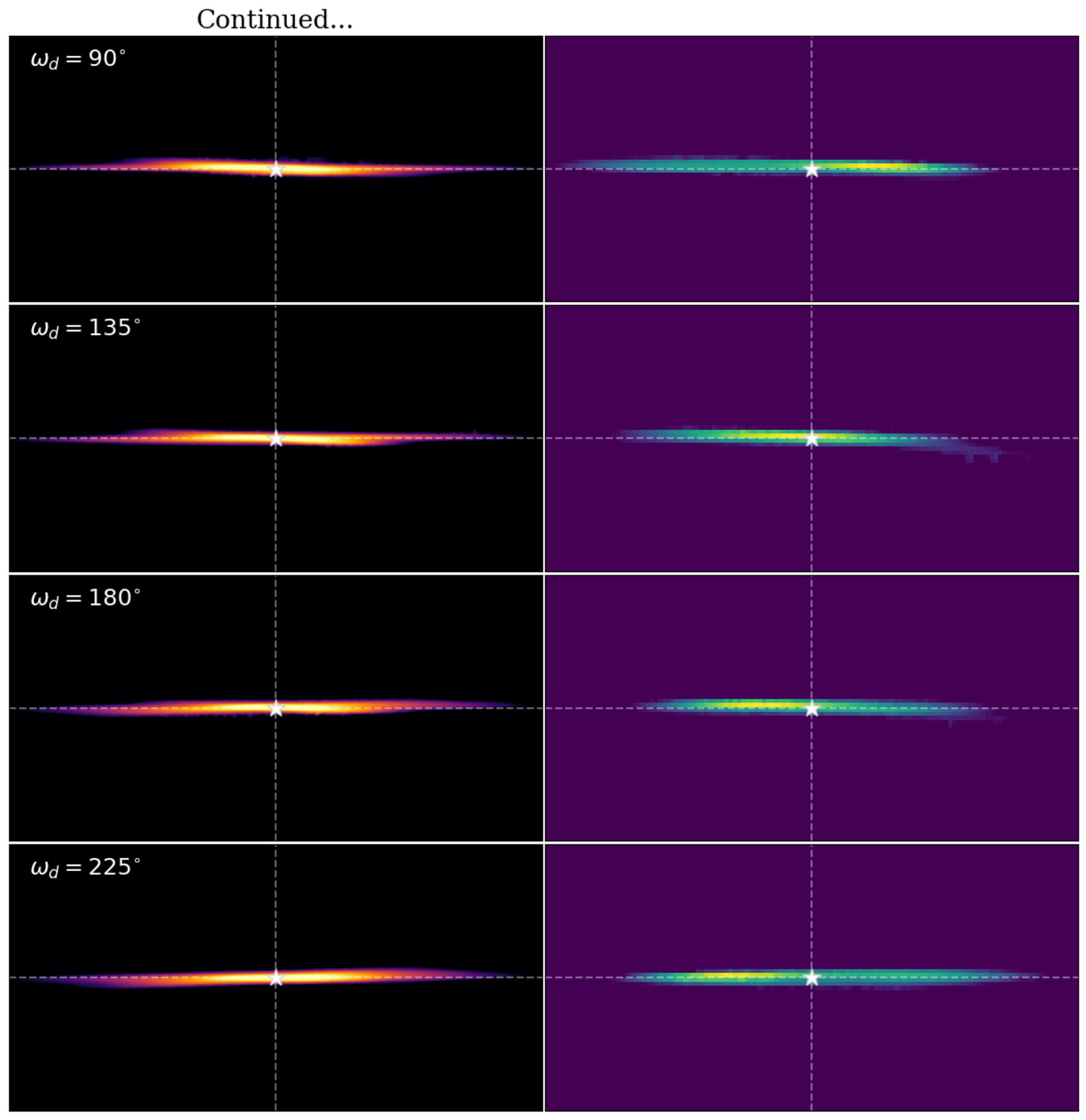}
\end{figure*}

\subsection{Planet Outside Warp vs. Inside Disk Inner Edge}
First, to understand what is happening in our simulations, Figure \ref{fig:density_1pl} shows the relative density of both disk models which are orientated face-on. For the 250pl model, because the planet begins embedded in the disk, the planet effectively carves a gap in the disk due to the chaotic unstable zone surrounding the planet \citep{Wisdom80}. This separates the disk into an inner and outer ring (which we refer to simply as the ``inner ring" and ``outer ring"), where the inner ring is significantly more densely populated than the outer ring. While the planet induces a visible eccentricity on the outer ring ($e \approx 0.3$), the inner ring remains relatively circular ($e \approx 0.03$) and appears axisymmetric. It is not entirely clear what the reason is for the significant difference in eccentricity between the inner and outer rings, although there may be several factors at play. For example, studies have shown that external perturbers can significantly deplete the outer edges of a debris disk, while the inner edges remain mostly unaffected. In comparison, internal perturbers are able to significantly deplete the inner and outer edges of an external debris disk \citep{Nesvold16}. This may help to explain the difference between the inner and outer ring morphologies, and why there appears to be no offset of the inner ring from the star.

For the 40pl model, the planet begins inside the disk inner edge and therefore does not carve a gap, although it does induce an eccentricity on the surrounding disk. The side of the disk near the planet's pericenter is closer to the star, while an over-density can be seen at the disk's apocenter. This is due to the fact that dust grains spend more time at apocenter, causing a pileup of particles at that location. Beyond 300 au, the 40pl model is fairly symmetric.

Comparing our inclined, synthetic scattered light models to HST/STIS observations of the disk halo as shown in Figure \ref{fig:1_planet_mods}, both scenarios are able to reproduce certain observed features including a warp due to the relative inclination of the planet to the disk. The angle of the warp in both cases is measured to be $\sim$3.3$^{\circ}$, which is smaller than the observed $4^{\circ}$, meaning that the relative inclination of the planet is likely to be close to, but slightly greater than 2$^{\circ}$. Both models also are able to produce a ``fork"-like bifurcated structure. The source of the fork feature is due to the planet exciting the inclination of nearby disk particles, causing these disk particles to oscillate about the planet's inclined orbit by $\sim$2$i_{pl}$ and create a second plane of the disk that extends from the warp location. The rest of the disk particles not affected by the planet remain aligned with the disk midplane creating the second half of the fork.

It is immediately apparent that the 250pl model is a better representation of the HD 111520 disk compared to the 40pl model. For example, the top and bottom sections of the fork in the 250pl model extend to the same distance on the right side of the disk, while only the bottom fork component is strongly seen on the left side, similar to the HST/STIS observations. The top part of the fork is still present on the left side, however, it is not as extended or bright as the bottom fork component which is aligned with the warp induced by the planet. In the 40pl model, the fork is not strongly seen, and the top and bottom sections of the fork do not extend to the same distance on either side. In fact, the part of the fork that aligns with the planet inclination and warp, does not extend past 300 au. In addition to the mismatch of the fork, the 40pl model also does not exhibit the radial extent asymmetry observed. Both sides of the disk, which are aligned with the midplane, appear to extend similarly out to $\sim$650 au, likely due to the planet not being able to induce any eccentricity at these distances by 15 Myr. The same cannot be said about the 250pl model, where the planet has strong influence on the outer ring causing a clear radial extent asymmetry, where the right side extends out to $\sim$600 au and the left side extends only to $\sim$400 au. Neither model appears to have a strong brightness asymmetry, although the 40pl model exhibits a modest pericenter glow on the left side of the disk caused by the planet's eccentric orbit. However, this is opposite of what is observed in the HD 111520 disk. 

Based on these two models, the 250pl model does a better job of recreating the majority of the observed disk features including the fork, warp, and radial extent asymmetry. The only feature that it is not able to reproduce is the strong brightness asymmetry. However, this is based solely on one orientation of the disk, i.e when the argument of pericenter of the disk ($\omega_{d}$) and the planet are $270^{\circ}$. Changing $\omega$ of the system by rotating the model counter-clockwise, may reveal a model that better matches the observed disk. Additionally, we can compare what our models might look like in the near infrared (NIR) with the GPI observations. To do this, we zoom in on the disk within 200 au, where the micron sized particles of the disk are located as observed by GPI. We then isolate the larger disk particles with $\beta < 0.2$, which are more concentrated close to the star in our models (see Section \ref{add_figs} located in the Appendix), in order to simulate the difference in dust grain sizes between the STIS and GPI observations. Similar to before, the model is inclined by $89^{\circ}$, binned into 4 by 4 au bins, and smoothed with a Gaussian kernel. 

Figures \ref{fig:250_rot} and \ref{fig:40_rot} show our two models rotated counter-clockwise by $45^{\circ}$ intervals starting with the initial $\omega_{d}$ of $270^{\circ}$.  Additionally, we keep the disk inclination at $89^{\circ}$. The left hand column for both figures show all the disk particles, which are highly dominated by the smallest particles, plotted within 650 au representing the disk as seen in the optical with STIS. The right hand column shows only the particles with $\beta < 0.2$ (particle radius $\gtrsim$ 5 $\mu$m), which are more concentrated in the inner regions of the disk, representing the disk as seen in the NIR with GPI. We find that the 40pl model cannot recreate the majority of disk features regardless of its orientation. While a radial and brightness asymmetry are present in the larger grains at certain orientations, the brightness asymmetry is still inconsistent, i.e. the side of the disk that is less radially extended is brighter. While this is expected for an eccentric disk, it opposite of what is observed in the HD 111520 disk where the more radially extended side is the brighter side. In the optical, the 40pl model also continues to lack a radial extent asymmetry or a similar fork structure. Due to these inconsistencies, we eliminate the 40pl model as a possible scenario for the HD 111520 system.

Taking a look at the 250pl models in Figure \ref{fig:250_rot}, we determine that the disk and planet $\omega$ are likely to be somewhere between $\sim$225$^{\circ}$ and $\sim$315$^{\circ}$. In these models, the disk retains a similar radial extent asymmetry and fork-like structure as observed in the optical. However, despite changing the orientation, no brightness asymmetry is observed. This is also true for the disk in the NIR, where no brightness or radial asymmetry are present, likely due to the fact that the inner ring stays symmetric as seen in Figure \ref{fig:density_1pl}. Therefore, while a planet on an eccentric, inclined orbit outside the warp location can easily create a similar warp, radial asymmetry and fork-like structure, another explanation is needed for the strong brightness asymmetry observed.

\begin{figure*}[ht!]
    \centering
    \caption{\label{fig:2pl_init} Initial (\textbf{left}) and final (\textbf{right}) orbits (black ellipses) and positions of the planets (black dots) relative to the disk (colored dots) for our 2 planet model. The two planets start at their pericenters with $\omega_{pl} = 270^{\circ}$ for the outer planet and $\omega_{pl} = 90^{\circ}$ for the inner planet. After 15 Myr, the outer planet is near its apocenter with $\omega_{pl} \approx 290^{\circ}$, while the inner planet is slightly past its pericenter with $\omega_{pl} \approx 180^{\circ}$.}
    \includegraphics[width=0.95\textwidth]{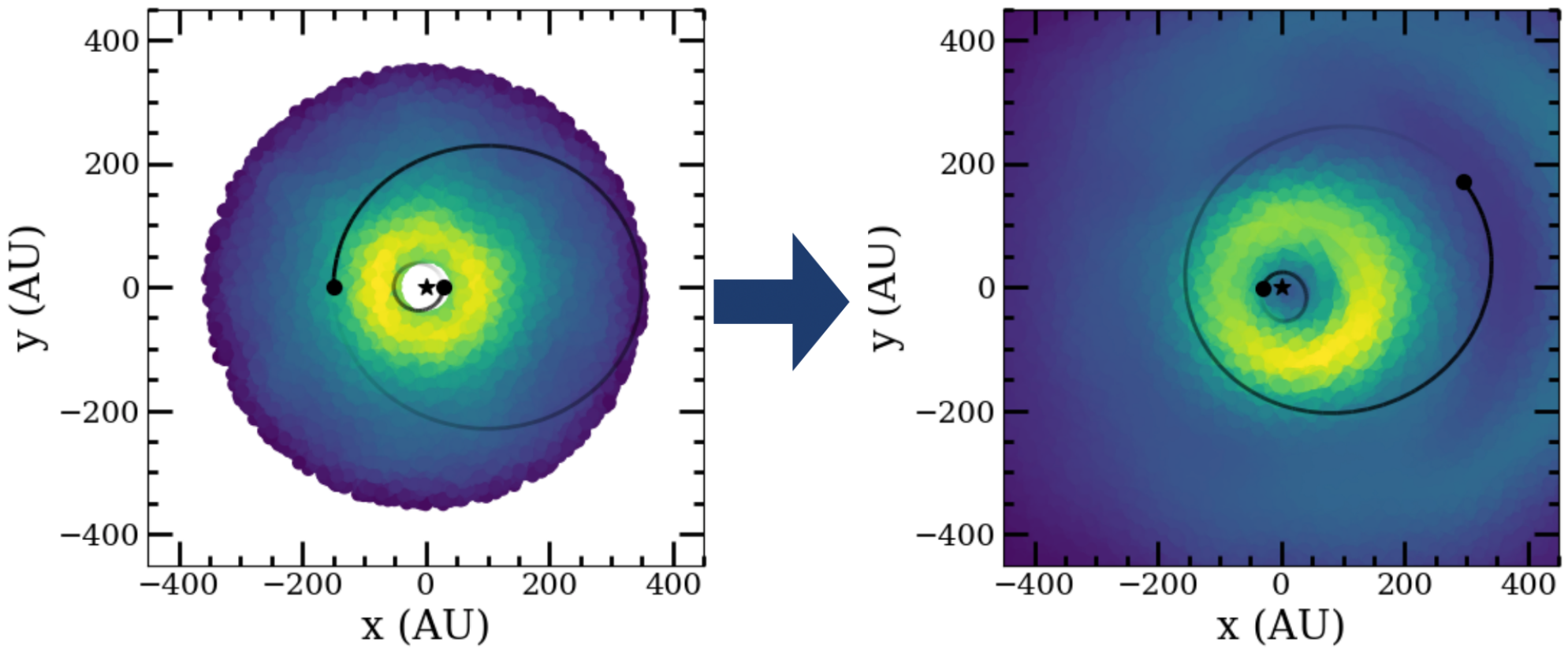}
\end{figure*}

\subsection{2 Planet Scenario}
Although the 250pl model is able to replicate very well the disk structure observed in the optical, the NIR model does not present a radial or brightness asymmetry. On the other hand, the 40pl is able to create a radial and brightness asymmetry, although the brightness asymmetry is on the wrong side as observed. We therefore combine these two models in attempt to replicate the structure seen in the disk halo as well as the brightness and radial asymmetry seen in the NIR by adding a second eccentric planet within the inner edge if the disk. By placing an eccentric planet within the disk inner edge, we can induce long lived eccentric structures on the inner disk (e.g. \citealt{faramaz14,naoz17}). Such an inner eccentric disk is supported by the GPI observations, which show that the polarized intensity peaks closer to star on the NW side than the SE side by 11 au. This result is opposite of what we observe in the disk halo, where the radial extent asymmetry suggests that the disk pericenter is on the SE side rather than the NW side. Such a discrepancy may be mitigated by the presence of an inner and outer disk component as found in 250pl model, where the inner disk component has a pericenter on the opposite side compared to the outer disk component. Additionally, if the pericenter of the inner disk is on the NW side, it may be possible to create the observed brightness asymmetry. 

For our 2 planet model setup, we keep exactly the same planet as in the 250pl model, and add an additional planet inside the disk inner edge similar to the 40pl model. We keep the inner planet at a semi-major axis of 40 au, but decrease the mass to Saturn size and decrease the eccentricity to 0.3 as a starting point. We also keep its orbit co-planar to the disk and define $\omega_{pl}$ to be $90^{\circ}$ so that the inner planet's pericenter is on the right side relative to the observer. Our reasoning for these changes is to attempt to create a pericenter glow on the right side of the disk as observed without significantly warping or disrupting the inner ring. The initial planet orbits and locations, as well as the location of the disk, can be seen in Figure \ref{fig:2pl_init}.

With our setup described above, we find that the resulting model is similar to the 250pl model, however, we are still not able to reproduce the radial or brightness asymmetry when inclining the disk without rotating $\omega_{d}$. This is surprising as the inner planet should induce some eccentricity on the inner ring. Taking a closer look reveals that due to secular perturbations from the outer planet, the argument of pericenter of the inner planet has shifted counter-clockwise from $90^{\circ}$ to almost $180^{\circ}$ by the end of the 15 Myr. Such a change may be due to the fact that anti-aligned planetary systems (where $\Delta \omega = 180^{\circ}$) can be unstable, especially for planets with high eccentricities that are not in mean-motion resonances (e.g. \citealt{Zhou03}). The $\omega_{pl}$ of the outer planet also shifts, but not as significantly from $270^{\circ}$ to $\sim$$290^{\circ}$. No major changes in $\omega_{pl}$ are seen in the previous single planet models.

\begin{figure}
    \centering
    \caption{\label{fig:2pl_rot} Synthetic scattered light image of the HD 111520 disk with 2 planets, orbiting outside and inside the warp ($a_{pl}=250$ and $a_{pl}=40$ au, respectfully). The disk is rotated counter-clockwise by 20$^{\circ}$ intervals from $\omega_{d} = 220^{\circ}$ to $\omega_{d} = 280^{\circ}$ \textbf{Left:} Representation of the disk as observed in the optical with HST/STIS \textbf{Right:} Representation of the disk as observed in the NIR with GPI. The rectangle in the top left figure represents the field of view of the NIR models.}
    \includegraphics[width=0.47\textwidth]{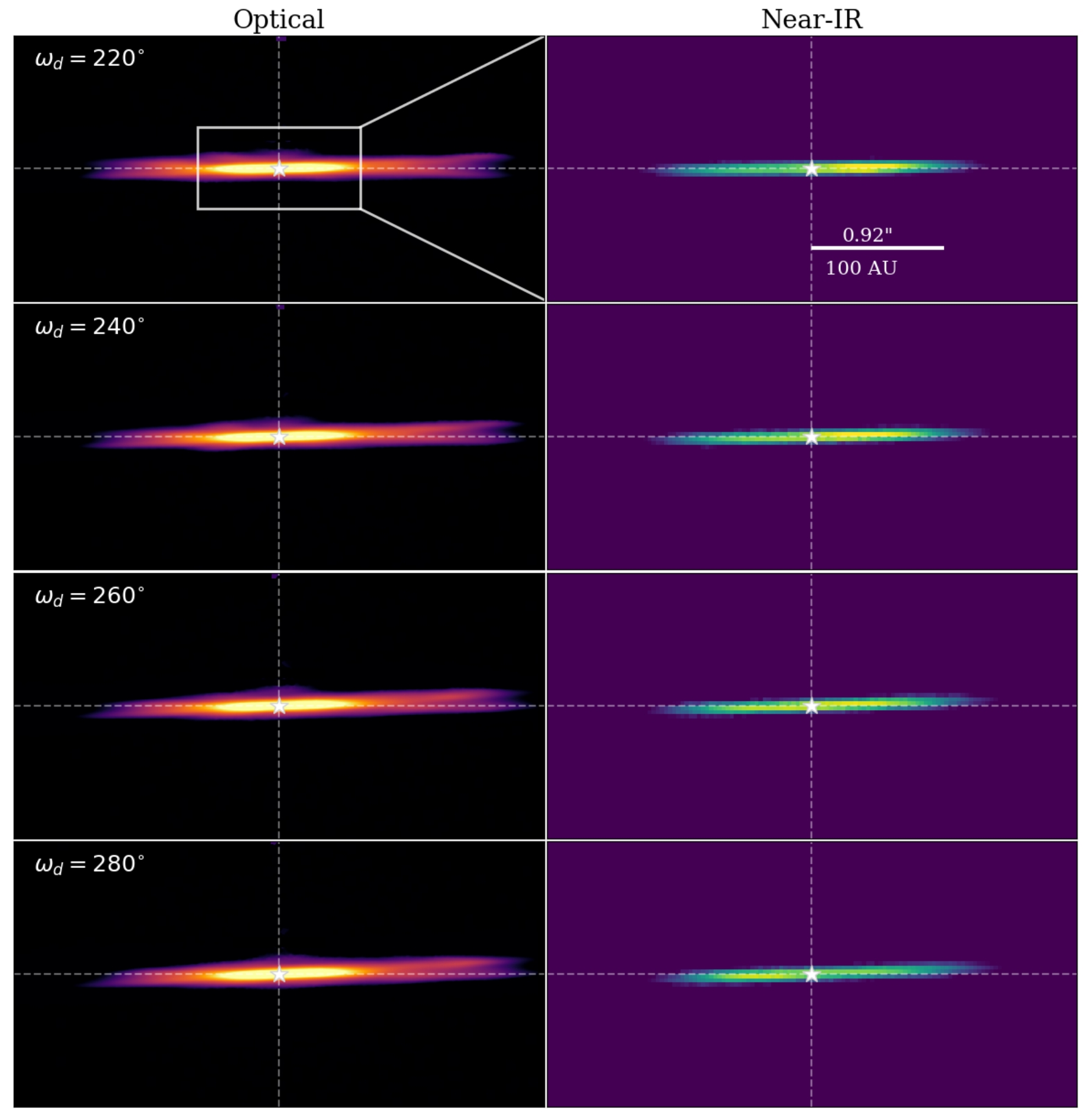}
\end{figure}

\begin{figure*}[ht!]
    \centering
    \caption{\label{fig:2pl_250} Synthetic scattered light image of the HD 111520 disk with 2 planets orbiting outside and inside the warp. The disk is rotated count-clockwise so that $\omega_{d} = 250^{\circ}$. The full disk is shown on larger scales, simulating STIS observations, while the larger disk particles are shown in the center, simulating GPI observations. The disk still exhibits the fork, warp and radial asymmetry, however, the inner planet has also created a brightness asymmetry on the NW side. The white dashed lines trace the ``fork"-like structure and warp induced by the outer planet.}
    \includegraphics[width=\textwidth]{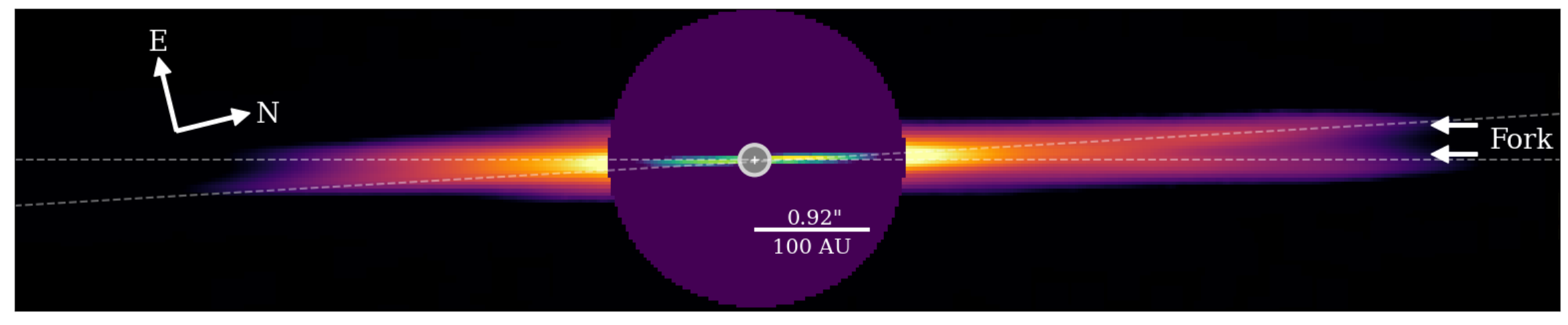}
\end{figure*}

Because $\omega_{pl}$ of the inner planet is close to $180^{\circ}$, the pericenter is facing away from the observer, and therefore no brightness asymmetry would be observed without rotating the disk's $\omega$ relative to the observer. We therefore rotate the model between $220^{\circ}$ and $280^{\circ}$ so that the fork and radial extent asymmetry are still visible and the pericenter of the inner planet is closer to the right side of the disk. As seen in Figure \ref{fig:2pl_rot}, we find that we are able to create a brightness asymmetry in the NW side with a $\omega_{d} < 260^{\circ}$, which also becomes more apparent the closer we get to $\omega_{d} = 220^{\circ}$. However, at the same time, we begin to slowly lose the radial extent asymmetry, and the top fork component on the SE side starts to become more visible. Figure \ref{fig:2pl_250} shows our model with $\omega_{d} = 250^{\circ}$, which is in between the middle two models shown in Figure \ref{fig:2pl_rot}. We choose to focus on this model as it still retains the fork, warp and radial asymmetry, but also harbours a brightness asymmetry similar to observations.

While our 2 planet model is able to replicate the majority of morphological features observed in the HD 111520 disk, there are still several caveats. In order to create the brightness asymmetry, the disk needed to be rotated so that $\omega_{d} < 260^{\circ}$ in order for the inner planet's pericenter to be closer to the NW side. This causes the top fork component on the SE side to become more visible, whereas it is not seen in observations, meaning that $\omega_{d}$ and $\omega_{pl}$ of the outer planet needs to be closer to $270^{\circ}$ when no brightness asymmetry is observed. Secondly, even when $\omega_{d} < 260^{\circ}$, the brightness asymmetry is not as significant as observed. For example, using our $\omega_{d} = 250^{\circ}$ model shown in Figure \ref{fig:2pl_250}, we measure the surface brightness over rectangular apertures, similar to what is done in previous studies \citep{Crotts22,Crotts24}. We find that the NW side is only 1.14 times brighter than the SE side in the NIR model, whereas the optical model shows no brightness asymmetry at the chosen aperture location (between 100 and 300 au, outside the HST coronagraphic mask). This is significantly lower than the 2 to 1 brightness asymmetry seen in GPI and the 5 to 1 brightness asymmetry seen with STIS. Even when rotating the disk so that $\omega_{pl}$ of the inner planet is at $90^{\circ}$, when we would expect the largest brightness asymmetry, only yields a 1.2:1 brightness asymmetry, again not close to the 2:1 asymmetry observed. 

It is important to note that our results for the 2 planet scenario are merely based on one configuration. It is possible with different orbital configurations and planet masses/eccentricities that we could achieve a larger brightness asymmetry more similar to observations, although this is outside the scope of our study. For now, we show that a planet near the warp location is able to create a similar fork, warp and radial asymmetry, while an additional inner planet is able to create a brightness asymmetry with a brighter NW side. While our model is not perfect, it is the first model that is able to replicate all of these features, making the HD 111520 system the perfect hunting ground for new directly imaged planets.

\section{Further Constrains on the Outer Planet Properties} \label{sec:constraints}
We presented three planet-disk models in attempt to explain the unique and complicated morphology of the HD 111520 debris disk. We find that a 1 M$_{jup}$ planet on an inclined and eccentric orbit with a semi-major axis of 250 au is better at recreating the disk morphology compared to the same planet with a semi-major axis of 40 au, i.e. inside the disk inner edge. We also find that at least one other planet is required to create a brightness asymmetry, although it is unclear whether this scenario is able to create a brightness asymmetry as extreme as observed. Despite this, our models show that planets can effectively create a similar complex morphology seen in the HD 111520 disk, and that planets may be responsible for other disks showing similar asymmetries and structures. Additionally, these models allow us to constrain the potential planet mass and orbit, which is important for understanding the feasibility of detection by current and future instruments. 

From our models, we know that the orbit of the outer planet is likely inclined by close to $2^{\circ}$ relative to the disk in order to produce the warp and fork, as well as likely eccentric in order to produce the radial asymmetry seen. We also know that a 1 M$_{jup}$ planet with a semi-major axis of 250 au is sufficient to replicate the disk halo structure. However, to understand the feasibility of detecting this planet, it is important to further constrain its potential mass and location. In the next few sections we take our 250pl model and vary the planet eccentricity, mass and semi-major axis with the goal of better constraining these properties.

\subsection{Planet Eccentricity} \label{sec:ecc}
\begin{figure*}
    \centering
    \caption{\label{fig:pl_ecc} Synthetic scattered light images of the HD 111520 disk with 1 planet orbiting outside the warp with varying eccentricity: $e = 0.1$ (\textbf{top}) and $e = 0.4$ (\textbf{bottom}). The star represents the location of the HD 111520 star, and both models are scaled similarly in log space.}
    \includegraphics[width=\textwidth]{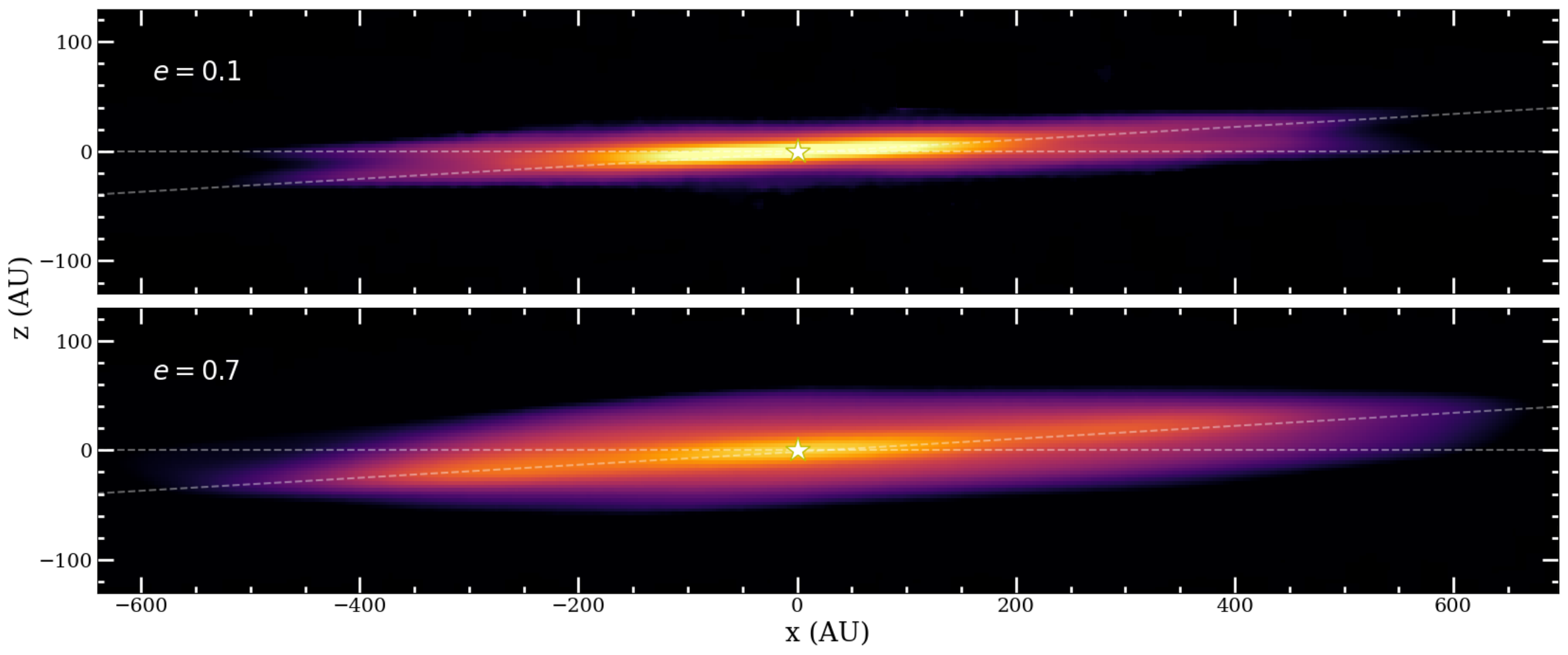}
\end{figure*}

As a starting point, we have set the eccentricity of the planet in our previous simulations to 0.4 in order to create the radial asymmetry observed with both STIS and GPI. While an eccentricity of 0.4 is successful in creating a radial asymmetry in the disk, here we test two other eccentricities to compare. Following the same procedure as before, keeping $m_{pl} = 1$ M$_{jup}$ and $a_{pl} = 250$ au, we simply change the eccentricity to 0.1 and 0.7. The resulting models can be seen in Figure \ref{fig:pl_ecc}. For simplicity we refer to these two models as the 0.1 and 0.7 models.

Comparing these two new models to the original model with $e = 0.4$ (250pl), show some similarities and some differences. Both models show warps where the disk deviates from the midplane, as well as a bifurcation of the disk to some degree. One difference between the two new models is that the fork structure is much more defined in the 0.1 model than the 0.7 model. This may be a result from the fact that the disk in the 0.7 model is significantly more puffy due the perturbations from such a highly eccentric planet. Additionally, because the pericenter of a planet with $e = 0.7$ is so close to the star, this completely disperses material within the planet's orbit. Both these factors of the 0.7 model are inconsistent with observations which suggest a more vertically flat disk with an inner radius of $\sim$50 au \citep{Draper16,Crotts22}. Therefore, a planet with a very high eccentricity seems unlikely.

While the 0.1 model more closely resembles the original 250pl model with $e = 0.4$ there are still a couple of key differences. For example, the top fork component on the left side of the disk in our model is more extended than either the original model or the 0.7 model, again inconsistent with observations. Additionally, the radial extent asymmetry between both sides of the disk is not as significant. Quantifying the radial extent asymmetry observed with STIS, we measure the surface brightness between 400 and 700 au on either side of the disk using several square apertures. Each aperture is 50 by 50 au large and is placed along the disk midplane, the flux is then averaged in each square aperture. Finally we look to see where the surface brightness reaches the noise floor on either side. With this procedure we find there is about 170$\pm$25 au difference between the radial extent of the NW (right) side and the SE (left) side. Comparing with our original model and the 0.1 model using a similar procedure with the synthetic scattered light images, we find that 0.1 model only has a difference in radial extent between the two sides of 70 au as seen from edge on, while the 250pl model has a difference in radial extent between the two sides of $\sim$170 au. 

Given the the inconsistencies between the 0.1 and 0.7 models, our initial moderate eccentricity of 0.4 is the most consistent with observations. This moderate eccentricity is able to create a similar radial extent asymmetry without vertically puffing the disk too dramatically or dispersing the inner disk. For the rest of this work, we will continue to use an eccentricity of 0.4 for consistency.

\subsection{Planet Mass} \label{sec:mass}
We next test the mass of the outer planet to see whether or not planet masses below or above 1 M$_{jup}$ can also create a similar disk morphology. The current planet mass limit set by GPI is $\sim$3 M$_{jup}$ for separations $>$ 10 au \citep{Nielsen19}.

For simplicity and consistency, we use our 250pl model with the single planet orbiting with a semi-major axis of 250 au. We keep the planet orbit exactly the same, only changing the planet mass. We test several different planet masses in comparison with the 1 M$_{jup}$, including 2 M$_{jup}$, 0.5 M$_{jup}$ and 1 Saturn mass ($\sim$0.3 M$_{jup}$). Each simulation is run exactly the same as before and the synthetic scattered light images are also produced the same way. Again, we focus mainly on whether or not the planet can produce the warp, fork and radial asymmetry as observed in the HD 111520 disk halo, ignoring the presence or absence of a brightness asymmetry. 

\begin{figure*}[ht]
    \centering
    \caption{\label{fig:pl_mass} Synthetic scattered light images of the HD 111520 disk with 1 planet orbiting outside the warp with varying masses: 2 M$_{jup}$ (\textbf{top}), 0.5 M$_{jup}$ (\textbf{middle}) and 0.3 M$_{jup}$ or 1 Saturn mass (\textbf{bottom}). The star represents the location of the HD 111520 star, and the center of the disk is masked to highlight the outer structure including the fork. As the planet mass decreases, the fork that is aligned with the midplane slowly starts to disappear on the NW side.}
    \includegraphics[width=\textwidth]{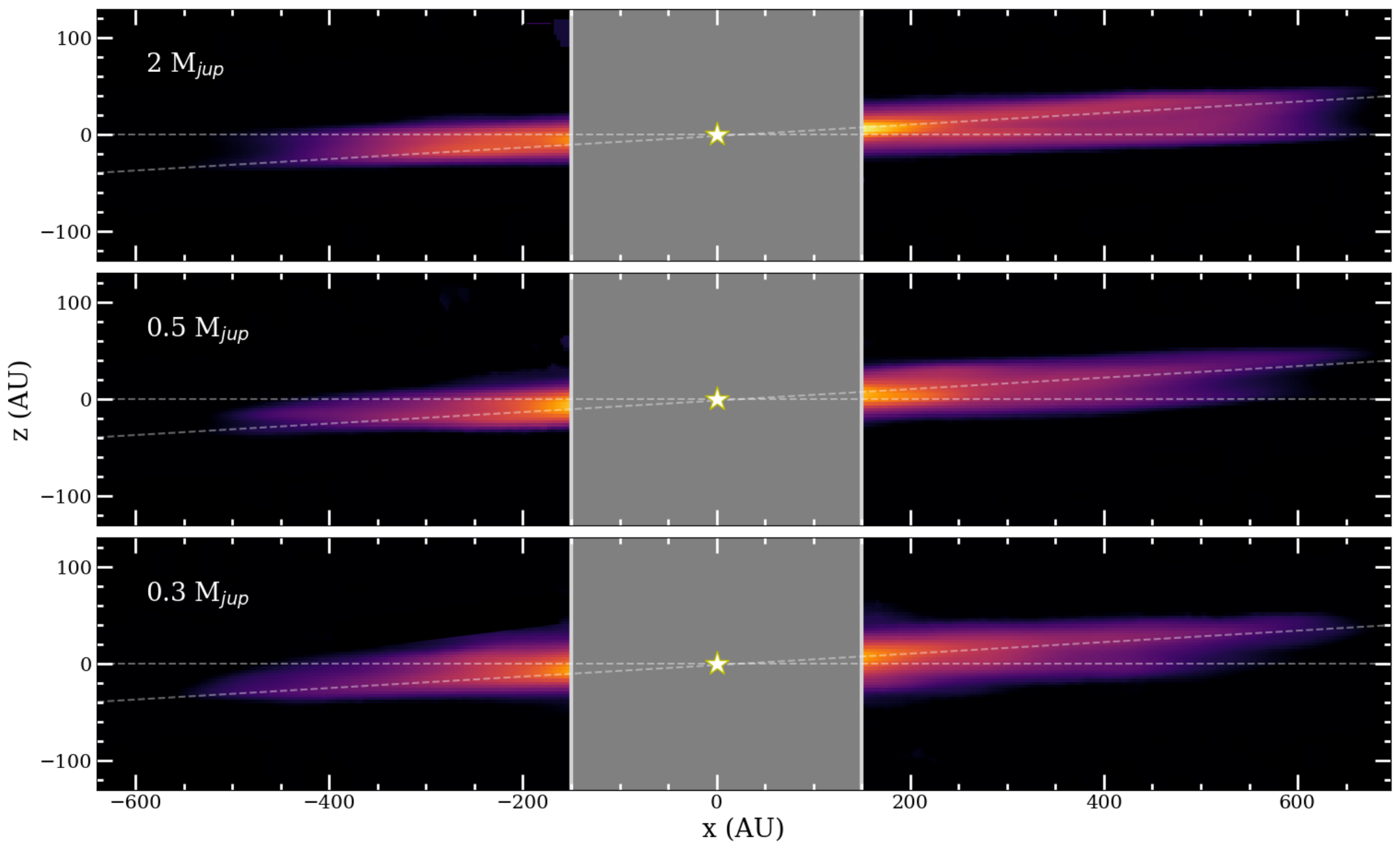}
\end{figure*}

The resulting models can be found in Figure \ref{fig:pl_mass}, where similar to previous figures, the model is orientated with $\omega_{d} = 270^{\circ}$. The center of the disk for each model is masked in order to solely focus on the outer structures of the disk. We find that the 2 M$_{jup}$ model is very similar to the 1 M$_{jup}$ model, in which the top and bottom fork components on the right side of the disk, as orientated in Figure \ref{fig:pl_mass}, radially extend to the same distance. Transitioning to lower planet masses, a clear trend emerges. Comparing the different models, the bottom fork component on the NW side that is aligned with the midplane appears to change, becoming fainter in the 0.5 M$_{jup}$ and 1 Saturn mass models. This result is unexpected and counter-intuitive, as one might expect the top fork component, which is directly influenced by the planet, to be the one affected. One possible explanation, is that the bottom fork component in the NW is correlated with the number of disk particles the planet is able to clear within the gap, given that the lower mass planets are unable to carve as deep of a gap within the same time span. Indeed, when comparing the number of disk particles in the gap versus the outer ring on the more radially extended NW side of the disk near the midplane ($z < 20$ au) for the 1 and 0.5 Jupiter mass planet models, we find that there are $\sim$1.5 times more particles in the outer ring compared to the gap for the 1 Jupiter mass planet model, where the opposite is true for the 0.5 Jupiter mass planet model (see Figure \ref{fig:1_0.5_mass} located in the Appendix). Additionally, there are overall $\sim$1.5 times more particles near the midplane in the outer ring of the 1 Jupiter mass planet model compared to the 0.5 Jupiter mass planet model. Similar analysis of the top fork component on the NW side of the disk reveals no significant difference in the number of particles between the two models. In summary, the planet mass does not significantly change the number of disk particles excited onto inclined orbits, but does correlate with the fraction of disk particles pushed out from the gap into the outer ring aligned with the disk midplane.  

In order to compare which model best represents the data, we attempt to quantify the difference in brightness between the top and bottom fork components on the NW side. We first measure the actual difference in brightness and uncertainty between the top and bottom fork component using the HST/STIS observations. Similar to measuring the brightness asymmetry between the NW and SE sides of the disk, we place two rectangular apertures centered on the top and bottom fork components between 450 and 550 au from the star, close to the edge of where the fork is still visible in the data, with a height of 14 au. Once the apertures are placed, the flux is integrated over the aperture and summed. We then divide the summed flux of the bottom fork component by the top fork component. To measure the uncertainty, we use the same two rectangular apertures and place them well above and below the disk emission, but at the same separation from the star (450 to 550 au). Again the flux is integrated over the aperture and summed, where the uncertainties in flux are then propagated to measure the uncertainty in the difference in flux between the top and bottom fork components. Using this procedure, we find that the top fork component is roughly 1.02$\pm$0.28 times brighter than the bottom fork component at this radial separation, meaning that both the top and bottom fork components are of similar brightness.

We conduct the same analysis for our various planet mass models, including the 1 M$_{jup}$, using the same rectangular apertures. From 2 M$_{jup}$ to 0.3 M$_{jup}$, the brightness of the bottom fork component compared to the top fork component consistently decreases. For the 2 and 1 M$_{jup}$ planet models we find the top fork component to be marginally brighter than the bottom fork component, where the top fork component is 1.24 and 1.26 times brighter than the bottom fork component, respectfully. As the planet mass decreases, the brightness asymmetry increases to 1.70 for the 0.5 M$_{jup}$ and 2.7 for the 0.3 M$_{jup}$ planet models. Based on these values, the 1 and 2 M$_{jup}$ models are the most consistent with observations within 1$\sigma$ uncertainties, while the models with a 0.5 Jupiter mass and 1 Saturn mass planet produce a bottom fork component that is much too faint. Given these measurements, we therefore put a lower mass limit on the outer planet of $\sim$1 M$_{jup}$ at a semi-major axis of 250 au. 

\subsection{Planet Semi-Major Axis}
Observations of the disk halo with STIS show that the disk is initially aligned with the midplane, before warping by $4^{\circ}$ from the midplane at $\sim$180 au from the star, while the fork feature becomes prominent beyond $\sim$245 au \citep{Crotts22}. We therefore test where these structures are seen in our REBOUND models, and compare how they change with the planet semi-major axis. To start, we test how a 1 Jupiter mass planet affects the disk morphology at varying planet semi-major axis ($a_{pl}$). We compare the final disk model when $a_{pl}$=150 au, 200 au, 250 au, and 300 au. We note that we are not trying to perfectly replicate the disk, but rather place some constraints on the possible orbit of the outer planet.

\begin{figure}
    \centering
    \caption{\label{fig:vert_prof_model} \textbf{Top:} Vertical offset profile for the 250pl model. The dark blue line represents the vertical offset measured using a single Gaussian, while the orange lines represent the vertical offset profile measured using a double Gaussian to highlight the fork structure. \textbf{Bottom:} Vertical offset profile of the GPI and HST/STIS observations of the HD 111520 disk, modified from Figure 5 in \citet{Crotts22}. The grey vertical lines represents the relative location where the top fork component converges with the bottom fork component in our model ($\sim$175 au). The dashed horizontal grey line highlights an offset of 0.}
    \includegraphics[width=0.47\textwidth]{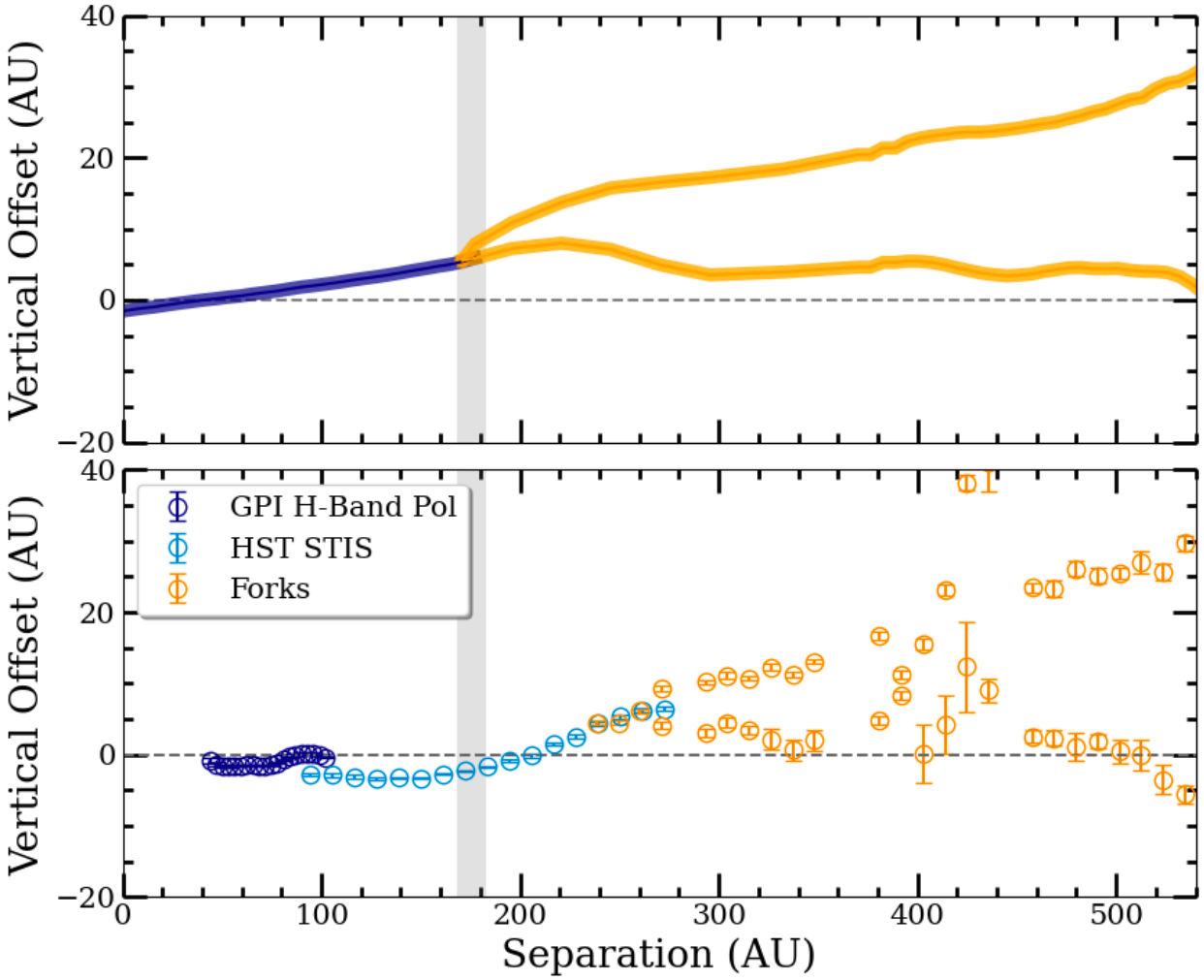}
\end{figure}

To determine the location of these structures, we use a similar method to \citet{Crotts22} by fitting a Gaussian profile to the disk surface brightness along vertical slices of the disk at varying radial separations, i.e. measuring the vertical offset of the disk from the star. We focus primarily on the NW side of the disk where the fork structure is detected. Between 0 and 240 au from the star, we fit a single Gaussian profile to our synthetic scattered light models, while we also fit a double Gaussian profile beyond 150 au to capture the location of the fork. Figure \ref{fig:vert_prof_model} shows the measured vertical offset for the 250 au planet model as an example.

When $a_{pl} = 150$ au, i.e. the planet is orbiting inside the warp location, we find that the inner ring almost disperses completely, similar to when the planet is highly eccentric, leaving only the disk component outside the orbit of the planet which is at large separations from the star. This is again inconsistent with observations with GPI, where the inner disk radius is estimated to be $\sim$50 au. Additionally, the lack of an inner ring results in similar issues as the 40pl model, where due to the planet's pericenter on the left (SE) side of the disk, a pericenter glow is observed in scattered light on the left side instead of the right side as observed. The fork also becomes less defined compared to the other models, making it difficult to fit a double Gaussian profile.

For the remaining three models, we compare the structure of the disk to observations. We find with our models that it is difficult to pinpoint where exactly the warp begins, as it is not as distinct in our vertical offset profiles compared to observations. We therefore focus on the structure of the fork. Comparing the location where the top and bottom fork components converge, we find that in all cases, the fork begins at roughly the same location ($\sim$175$\pm$5 au) regardless of the planet semi-major axis. The convergence of the top and bottom fork components in all three cases also appear to coincide with the outer edges of the inner ring, suggesting that the location of the inner ring may determine how far we can observationally probe the fork structure as seen edge-on in scattered light, given that the inner ring is much brighter than the fork. To investigate this correlation further, we look more closely at how the semi-major axis affects the inner ring by measuring its radial density profile.

\begin{figure*}
    \centering
    \caption{\label{fig:pl_sma} Dust density images of our disk model seen face on, using a 1 M$_{jup}$ planet with various semi-major axes (from left to right: $a_{pl} = 150$ au, 200 au, 250 au, and 300 au). The white dashed line in each plot highlights the location of the peak density of the inner ring for the 200, 250, and 300 au models, while the orange shaded line highlights the location where the top and bottom fork components converge on right side for the same three models ($\sim$175 au). The density is scaled to have similar brightness between each frame.}
    \includegraphics[width=\textwidth]{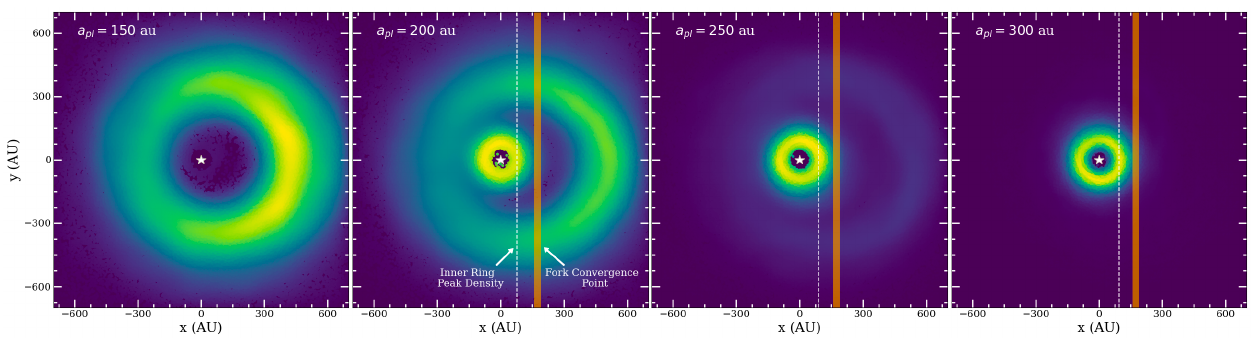}
\end{figure*}

Figure \ref{fig:pl_sma} shows the resulting density of the disk for each model with the disk oriented face on. We find that there are three properties of the inner ring that change with planet semi-major axis: The ring's density; location of the peak density; and its FWHM. Going from a smaller semi-major axis to a larger semi-major axis, the inner ring becomes more dense as less disk particles in this region are swept out by the planet, while the outer ring becomes relatively less dense. At the same time, the radius of the peak density of the inner ring increases slightly from $\sim$76 au to $\sim$95 au, while the FWHM of the inner ring decreases from $\sim$114 au to $\sim$90 au. This can be seen in Figure \ref{fig:pl_sma}, where the peak density of the inner ring appears to move outward with $a_{pl}$, while also decreasing in width. One explanation for these correlations is that as $a_{pl}$ decreases, the inner ring becomes more truncated as the chaotic zone around the planet moves in, also resulting in the removal of more disk particles closer to the star. This would explain why the peak density of the inner ring moves in, and why there are less particles in the inner ring as the planet's orbit decreases. The increase in the FWHM as $a_{pl}$ decreases may be due to increased stirring of particles in the inner ring, where a wider disk indicates a higher level of stirring due to an increased distribution of particle eccentricities \citep{AM09}. Due to the relationship between the inner ring location and FWHM with increasing $a_{pl}$, the outside of the inner ring remains at relatively the same radial distance from the star. This means that the fork, as viewed in scattered light from edge on, becomes distinct at the same location for all three models.

Regardless of the differences between the three models, for all three cases, the fork structure begins closer to the star than observed with STIS by $\sim$60 au. However, it is unclear if this difference is real or due to our model setup or optical affects making it difficult to probe the fork fully in the observations. In either case, the main importance is that the fork in our models ($a_{pl}$=200, 250 and 300 au) is present at the location seen in the data ($>$245 au) while not compromising the rest of the disk morphology. We therefore place a lower limit on the planet semi-major axis at approximately 200 au for a 1 Jupiter mass planet.   

\subsection{Combining Planet Mass and Semi-Major Axis}
\begin{figure*}
    \centering
    \caption{\label{fig:pl_sma_mass} Table of models with varying planet mass and semi-major axis. Dark red filled cells represent models that are inconsistent with observations, orange cells represent models that are mostly consistent with observations, and green filled cells represent models that are consistent with observations. Each one of these cells list the surface brightness ratio between the top and bottom fork components on the right side of the disk, if an inner ring is present or not, and the radial extent asymmetry between the left and right sides of the disk. Light red filled cells represent parameter space that is inconsistent with observations based on extrapolations from the models produced in the dark red cells.}
    \includegraphics[width=\textwidth]{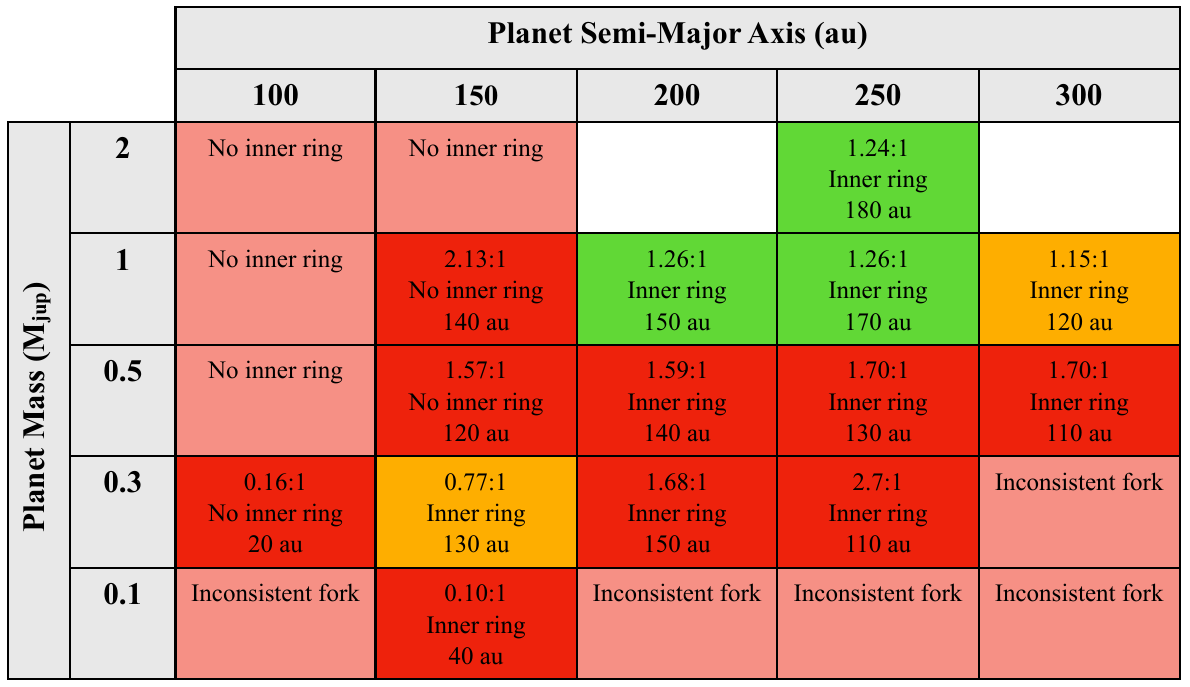}
\end{figure*}

In the previous two sections, we place lower limits on the planet mass at 1 Jupiter mass at a semi-major axis of 250 au, and place a lower limit on the semi-major axis of 200 au for a 1 Jupiter mass planet. However, to get a better sense on the lower limits for these two parameters, the mass and semi-major axis should be varied together rather than separately. We therefore run several additional simulations using the same procedure as before, testing planet masses from 0.1 to 2 Jupiter mass and a semi-major axis from 100 to 300 au. Based on our previous findings, we look for three particular features to determine the consistency of our models with observations; 1) Is there an inner ring? and 2) Are the top and bottom fork components on the right side of the disk of similar brightness (within the 1$\sigma$ uncertainty measured in Section \ref{sec:mass})? and 3) Is there a significant radial extent asymmetry? The results of our simulations can be found in Figure \ref{fig:pl_sma_mass}.

In Figure \ref{fig:pl_sma_mass}, we present a table of planet mass versus semi-major axis ranging from 0.1 to 2 M$_{jup}$ and 100 to 300 au. Each cell represents a different scenario, where the dark red, orange, and green cells represent scenarios in which we run a simulation and create a synthetic scattered light model. For the scenarios in which we create a model, we measure the ratio of the surface brightness between the top and bottom fork components (stated in each cell) using the same procedure as in Section \ref{sec:mass} and look for the presence of inner ring. Additionally, we measure the difference in the radial extent between the two sides using the same procedure as in Section \ref{sec:ecc}. The dark red cells are scenarios in which the model is not consistent with observations, either because the inner ring is completely dispersed and/or the top and bottom fork components on the right side of the disk are inconsistent in brightness. The orange cells represent scenarios where the model meets the first two criteria, but do not have a large enough radial extent asymmetry. Finally, the green cells are scenarios in which the model is consistent with observations across all three criteria. The light red cells are scenarios in which we do not run a simulation, but rather extrapolate our findings from the dark red cells in which a model is produced and found to be inconsistent.

Based on these results, we find that our lower limits on the planet mass and semi-major axis from the previous two sections has not changed significantly. Beyond a semi-major axis of 200 au, the lower limit on the planet mass remains $\sim$1 M$_{jup}$, as below this mass the bottom fork component on the right side of the disk becomes too faint. Within a semi-major axis of 150 au, things change slightly. As we go from a larger planet mass to a smaller planet mass, the top fork component instead of the bottom fork component becomes more faint as the planet is unable to perturb the outer regions of the disk similar to what occurs in the 40pl model. Only one model with a planet mass of 0.3 M$_{jup}$ and a semi-major axis of 150 au is able to replicate observations, where the planet is small enough that the inner disk is not completely dispersed, but big enough that the top and bottom fork components on the right side of the disk are of similar brightness. However, the radial extent asymmetry for this model is 40 au smaller than observations (130 au difference compared to 170 au). We therefore place this model in the orange category as the radial extent asymmetry is still within 2$\sigma$ uncertainties of observations and the radial asymmetry may vary slightly depending on the placement of the particles generated in the creation of the synthetic scattered light image. We also place the model with a planet mass of 1 M$_{jup}$ and a semi-major axis of 300 au in the orange category for similar reasons, however, the decrease in the radial extent asymmetry is more a result of our chosen initial outer disk radius of 350 au, meaning that the planet's apocenter starts outside of the disk. This causes the disk to become more truncated. Within 150 au, the semi-major axis is too small to replicate observations at any planet mass due to the complete dispersal of material within the planet's orbit in addition to the top fork component on the right side of the disk not being visible out to 600 au.

To summarize, for a small range of planet masses ($\sim$0.3-0.4 M$_{jup}$), it is possible for the outer planet to have a semi-major axis as small as 150 au, although the radial extent asymmetry is less significant in these cases. Beyond a semi-major axis of 200 au, however, the minimum mass of the outer planet remains around 1 M$_{jup}$.

\section{Discussion} \label{sec:discussion}
In the previous Section, we are able to use the disk morphology to place some constraints on the planet eccentricity, mass, and semi-major axis, based on how varying these parameters affects different disk structures. Here, we discuss the implications of our results, and what they might mean for planet detectability, as well as the evolution of the system as a whole. 

\subsection{Alternative Explanations}
With our n-body simulation setup, we demonstrate that the HD 111520 debris disk morphology can be mostly reproduced with a single Jupiter mass planet on an eccentric, inclined and wide orbit. However, one feature of the HD 111520 disk we are not able to fully replicate is the extreme brightness asymmetry observed in both the optical and NIR. While the addition of an inner planet with the right orbital configuration is able to create a brightness asymmetry without significantly altering the rest of the disk morphology, the resulting brightness asymmetry is not as strong as observed and is not observed in the disk halo. An extremely eccentric planet may be required to create a 2 to 1 brightness asymmetry, but such a planet is not supported by the data. For example, \citet{Crotts24} found that the eccentricity measured from the GPI data, either through the measured disk geometry or the 11 au difference in the peak polarized intensity as measured in \citet{Crotts22}, would not be able to reproduce as strong of a brightness asymmetry. It is therefore possible that another mechanism is required. 

One alternative scenario is a recent giant collision between two rocky bodies, which is discussed briefly in \citet{Crotts22} and also explored in depth by \citet{Jones23}. Although \citet{Jones23} find that a giant collision scenario is also able to reproduce a fork, radial and brightness asymmetry, their model has similar issues with our model, in that it is not able to properly reproduce the brightness asymmetry. This discrepancy is due to the fact that in order for the NW side to be bifurcated and more radially extended, the collision needs to take place on the SE side. However, this then leads to a brighter SE side, which is the opposite of what is seen in observations. Instead, it is possible both mechanisms are required to produce the structure of the disk while also producing the extreme brightness asymmetry. If the collision point location is instead located on the NW side, this may be able to explain the extreme brightness asymmetry, as well as the difference in disk color measured \citet{Crotts22} where the NW side is significantly more blue than the SE side at shorter wavelengths indicating a larger concentration of small dust grains (e.g. \citep{Boccaletti03}). Future work would be required to test whether or not this is a plausible theory for HD 111520, although a similar scenario can be found through the $\beta$ Pic disk, which harbours both a warp driving planet and a suspected recent giant collision \citep{Dent14}. Given the similarities between HD 111520 and $\beta$ Pic, the HD 111520 disk makes a good candidate for JWST/MIRI observations to search for similar structures in the disk emission such as the famous ``cat tail" seen in $\beta$ Pic which is thought to be one result of the giant collision \citep{Rebollido24}.

Another possible scenario is that a recent stellar flyby perturbed the HD 111520 debris disk. While stellar flybys are rare, they are more likely to occur in young stellar associations (such as the LCC group) due to a higher stellar density. For example, HD 106906 is another debris disk system in the LCC group which is suspected to have been affected by a recent stellar flyby as a way to explain the location/orbit of the planet HD 106906 b. A close stellar flyby may also affect the orbit of dust particles in a disk. In \citet{Rodet19}, the authors study the effect of a stellar flyby on the HD 106906 disk, and found that the flyby can increase the eccentricity and inclination of disk particles with the strongest effects at the outer reaches of the disk. This could explain why the warp and fork are present outside the planetesimal belt in the HD 111520 disk halo ($\gtrsim$180 au), whereas the planetesimal belt as observed with GPI remains vertically flat. However, it is unclear how long these perturbations can be sustained without continued perturbation, and whether or not this scenario could explain the complex structure of the disk as a whole. In fact, \citet{Rodet19} find that planet, HD 106906 b, would have a much greater affect on the disk than a stellar flyby, which would have to come within 0.05 pc (3000 au) of the disk. Additionally, even though the LCC has a higher stellar density, the possibility of a stellar flyby is still relatively rare. In the case of the HD 106906 system, \citet{derosa19} found only 2 possible stellar candidates for flybys out of 461 nearby stars ($< 1 \%$). While we cannot completely rule out a flyby scenario without further research, it still remains a less likely scenario compared to perturbations from a planet which can also create long-lived structures in the disk.

\subsection{Outer Planet Detectability}
If there is indeed a planet orbiting at wide separations in the HD 111520 system, the question then becomes whether or not we can detect such a planet with the constraints placed on the mass and orbit. Based on the resulting disk morphology, we placed a lower limit of $\sim$200 au (1.9$''$) for the planet's semi-major axis. At this distance, the planet is not detectable through methods such as transit or radial velocity which require the planet to be much closer to the star. Because HD 111520 is a young system (15 Myr) and the planet is at a wide separation, direct imaging would be the best method for detection as the planet would still be warm from formation. In this case, a high-contrast imaging instrument in the optical/NIR with a wide field of view (FOV) would be required.  

Four such instruments includes (but are not limited to) the STIS instrument on HST, GPI, the Spectro-Polarimetric High-contrast Exoplanet REsearch (SPHERE) on the Very Large Telescope, and the NIRCam instrument on JWST. In the case of HST/STIS, GPI, and VLT/SPHERE, all three instruments have already imaged the disk and have not detected any planets \citep{DP15,Nielsen19,Xie22}. In the case of HST/STIS, the disk halo almost completely coincides with the planet's orbit. Additionally, planets are not hot enough to emit significantly in the optical, therefore the NIR would be more ideal for detection. The lack of detection with GPI is likely due to its small FOV (2.8$''$ by 2.8$''$) while lack of detection with SPHERE is likely due to the mass of the planet, where even young 1-2 M$_{jup}$ planets tend to have a lower temperature compared to their more massive multi-Jupiter sized counterparts, therefore requiring a very deep contrast which is difficult for most current instruments to achieve. JWST/NIRCam therefore provides the best chance to observe this planet, as the instrument has a large FOV (10$''$ by 10$''$) and can achieve deeper contrasts than other high-contrast imagers.

To calculate whether or not we can observe the proposed outer planet with NIRCam, we compute planet mass sensitivity curves for the F444W filter ($\lambda_{c} = 4.44$ microns). We initially use the code panCAKE \citep{Carter21} to create 5$\sigma$ contrast curves, and find that between 200 au and 300 au ($\sim$1.9$''$-2.8$''$) we are able to achieve a contrast of between 1e-6 and 2e-6. Inputting these contrasts into a planet evolutionary code (ATMO; \citealt{Phillips20}), we find for a 15 Myr system, NIRCam is able to reach a planet mass of 1 M$_{jup}$ with a $\sim$75\% probability of detection at the given separations. 

Our calculations show that JWST/NIRCam in the F444W filter can detect lower mass planets than the lower limit established between 200 and 300 au by our simulations. Even so, there may still be some concerns with the planet's detectability. One possible concern is interference from the disk. Because we are considering NIR observations, the disk should not have a significant impact as the micron sized grains are more radially compact than the sub-micron sized grains as seen in the optical with HST. Additionally, because the planet's orbit is close to edge on, meaning that it may be located at small separations from the star within NIRCam's inner working angle ($\lesssim$60 au) depending on where the planet is in its orbit. While this is definitely possible, because the planet is likely eccentric given the radial extent asymmetry observed in the disk, it is statistically more likely to be near apocenter as it would spend the majority of time here (P $\approx$ 3523 years for a 1 M$_{jup}$ planet with $a_{p}$=250 au). Therefore, JWST provides an excellent opportunity to detect the warp driving planet predicted by our models.

\subsection{Implications on Planet Formation and Evolution}
In addition to assessing the planet's observability, the constraints placed on the planet's mass and orbit also provide information on its formation and evolution. Given that the planet is highly aligned with the disk (again, $\Delta i \approx 2^{\circ}$), it is most likely that the planet formed within the disk. One of the main questions that arises then is how did the planet end up on such a wide and eccentric orbit? One possibility is that the planet formed via disk fragmentation \citep{Toomre64,Goldreich65}, as this formation method is able to form wide orbit sub-stellar companions with a wide range of eccentricities. However, with a planet mass of $\sim$1 M$_{jup}$, formation through gravitational instability is less likely to be the case as this method is preferred for formation of more massive companions such as brown dwarfs (e.g. \citealt{Kratter16,Forgan13}). Additionally, while sub-stellar companions formed via disk fragmentation may start out aligned with the disk, studies show that over time these companions are likely to become misaligned due to interactions with other companions in the system \citep{Stamatellos09}. This is supported by the fact that brown dwarfs have been observed to have a high likelihood of misalignment from the star and a wide range of eccentricities \citep{Bowler20,Bowler23,Nagpal23}. On the other hand, the same studies also found that wide-orbit, directly imaged planets are much more likely to be aligned with their star and have a lower range of eccentricities.

Another possibility is that the planet formed via core accretion, a mechanism that prefers the formation of smaller planets compared to disk fragmentation \citep{Goldreich04}. While this mechanism is more likely to result in a 1 M$_{jup}$ planet that is coplanar, one issue with this scenario is that it does not explain the planet's wide and eccentric orbit. For example, planets observed between 10 and 100 au through the GPIES campaign tend to have orbits closer to 10 au compared with the observed brown dwarfs \citep{Nielsen19}. In this case, another mechanism is required to explain the high eccentricity and large . A possible explanation is that the planet at some point was scattered outward onto a high eccentricity orbit through dynamical interactions with another object. A similar scenario is theorized for the planet-disk system, HD 106906, where the directly imaged planet is thought to have scattered to its current position ($\sim$730 au from the star; \citealt{Bailey14}) due to interactions with the close central binary \citep{Rodet19}. In the case of HD 111520, while the system does have a stellar companion, it is widely separated ($\sim$17000 au) with a PA of $78^{\circ}$ \citep{BM13} meaning that it is unlikely to be the scattering culprit. Instead, our planet's orbit may point towards scattering with another planet in the system, a process which is thought to be common for giant exoplanets (e.g. \citealt{Chaterjee08,Bitsch20}). Therefore, the constrained orbit and mass of the outer planet in our models may further suggest the presence of at least one inner planet in the HD 111520 system. However, it is not clear whether or not a planet would be able to achieve a semi-major axis as large as 250 au with a planet-planet scattering scenario, and whether or not the debris disk would be fully disrupted in the process. To summarize, while it is likely the planet formed within the disk, further studies will be needed in order to investigate whether a disk fragmentation or core accretion and planet-planet scattering scenario is more likely.

\section{Conclusion}
In this study, we model the highly asymmetric debris disk around HD 111520 using the n-body simulation code REBOUND, with the goal of determining what kind of planet(s) can reproduce the disk morphology, if any. We find that $\sim$1 M$_{jup}$ planet on a wide and eccentric orbit that is also inclined by $\sim$2$^{\circ}$ relative to the disk, is able to produce a warp, ``fork"-like structure, and radial asymmetry. We compare two models where $a_{pl}$=40 au and 250 au, and find that the planet with $a_{pl}$=250 is much better at replicating the overall disk morphology and is likely to have an argument of pericenter close to $270^{\circ}$. While an inner planet may be required to create a brightness asymmetry between the NW and SE sides, we are unable to replicate the strength of the brightness asymmetry observed, suggesting that another mechanism (in addition to the outer planet) may be needed, such as a giant collision similar to the $\beta$ Pic disk.

In an attempt to further constrain the properties of the planet, we vary the outer planet's eccentricity mass, and semi-major axis to see how this affects the morphology of the disk. When comparing a low, moderate, and high planet eccentricity, we find that a moderate eccentricity ($e = 0.4$) is able to replicate observations the best as the planet is able to induce a sufficient radial extent asymmetry without significantly disrupting the disk. Varying the planet mass shows that as the mass decreases, the brightness of the bottom fork component on the NW side also decreases. Additionally, decreasing the planet semi-major axis can decrease the radial extent asymmetry and may also cause the inner disk to completely disperse creating an inconsistent disk morphology. By varying the planet mass and semi-major axis in tandem, we find that a planet with $\gtrsim$1 M$_{jup}$ and $a_{p} \gtrsim 200$ au is able to reproduce the overall disk morphology the best as it creates a comparable radial extent asymmetry and fork structure without disrupting the inner disk. These constrained parameters, alongside its mutual inclination with the disk, would suggest that the planet was formed within the disk, although it is not clear whether the planet was formed via disk fragmentation or core accretion and experienced a planet-planet scattering effect.

The HD 111520 debris disk is a unique system, and provides an excellent opportunity to study disk-planet interactions. We demonstrate how the complex morphology of the disk can be used to infer an unseen planet, as well as constrain certain planet properties such as the mass and orbit. This work also demonstrates how disk structures such as warps and ``forks" may be the best signposts for planets compared to other disk asymmetries such as brightness asymmetries. Finally, HD 111520 makes a great candidate for future observations with JWST NIRCam to search for our warp-driving planet and thereby add another system to the small list of directly imaged planet and well resolved disk systems. 

\begin{acknowledgements}
The authors wish to thank the anonymous referee for helpful suggestions that improved this manuscript. The authors also wish to thank Dr. Hanno Rein for their expertise and help regarding REBOUND and Dr. Gaspard Duch\^{e}ne for their insightful suggestions and feedback. KAC and BCM acknowledge a Discovery Grant from the Natural Science and Engineering Research Council of Canada. This work made use of data from the European Space Agency mission \emph{Gaia} (\url{https://www.cosmos.esa.int/gaia}), processed by the \emph{Gaia} Data Processing and Analysis Consortium (DPAC, \url{https://www.cosmos.esa.int/web/gaia/dpac/consortium}). Funding for the DPAC has been provided by national institutions, in particular the institutions participating in the Gaia Multilateral Agreement. This research made use of the SIMBAD and VizieR databases, operated at CDS, Strasbourg, France. 
\end{acknowledgements}

\clearpage

\appendix

\section{Additional Figures} \label{add_figs}

\subsection{Radial Distribution of Disk Particles vs. Particle Size}
\begin{figure*}
    \centering
    \caption{\label{fig:250pl_bet} Radial density distribution of massless disk particles with different $\beta$ values in our 250pl model. \textbf{Left:} Distribution of particles with $\beta < 0.1$. \textbf{Middle:} Distribution of particles with $0.1 < \beta < 0.3$. \textbf{Right:} Distribution of particles with $\beta > 0.3$. For all three frames the density is scaled the same in log space.}
    \includegraphics[width=\textwidth]{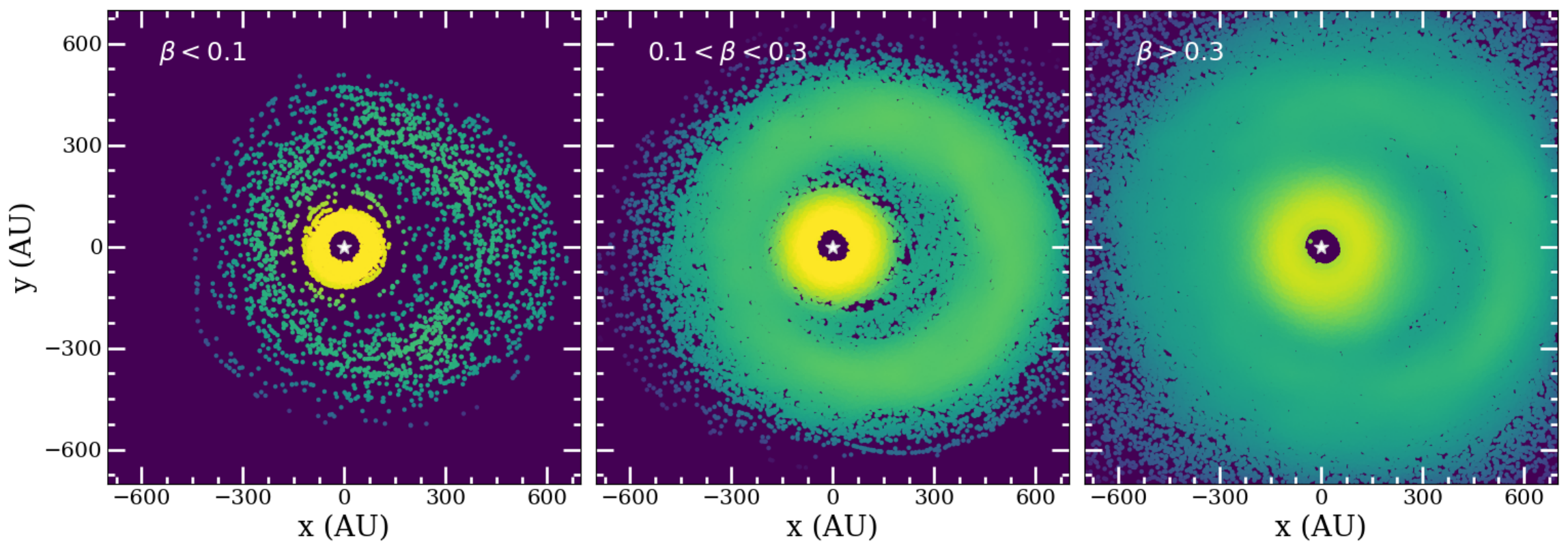}
\end{figure*}

\begin{figure*}
    \centering
    \caption{\label{fig:40pl_bet} Radial density distribution of massless disk particles with different $\beta$ values in our 40pl model. \textbf{Left:} Distribution of particles with $\beta < 0.1$. \textbf{Middle:} Distribution of particles with $0.1 < \beta < 0.3$. \textbf{Right:} Distribution of particles with $\beta > 0.3$. Again, for all three frames the density is scaled the same in log space.}
    \includegraphics[width=\textwidth]{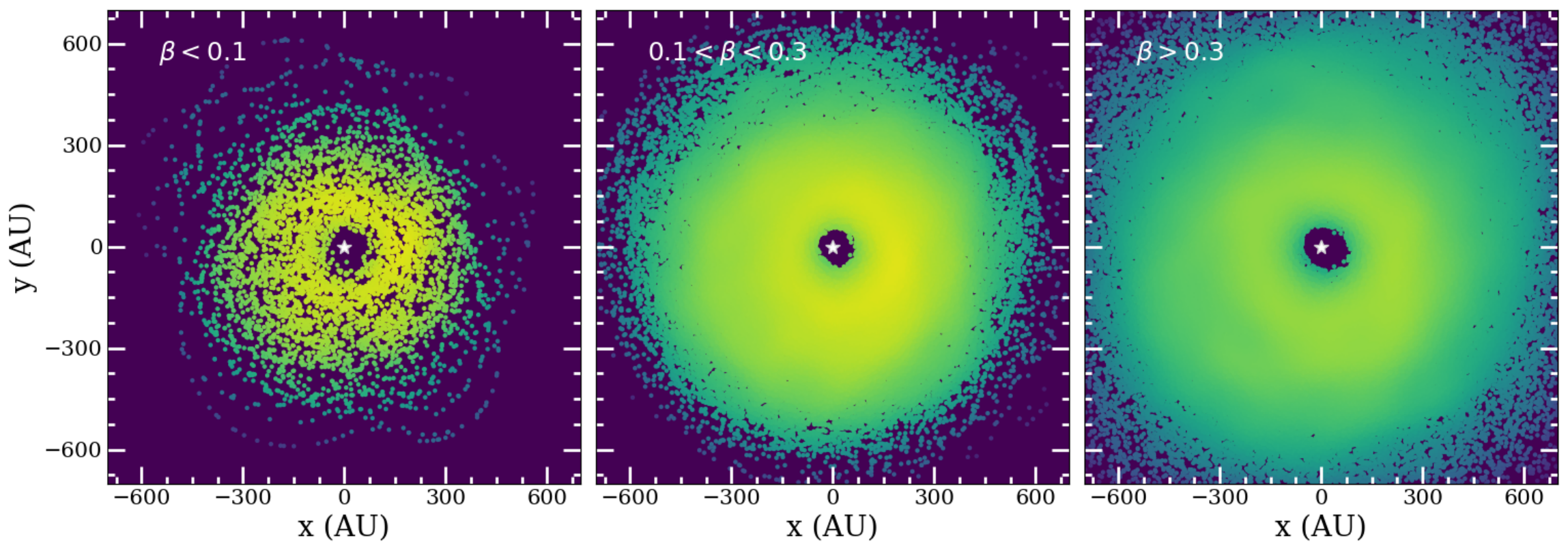}
\end{figure*}

In Figures \ref{fig:250pl_bet} and \ref{fig:40pl_bet}, we show the radial density distribution of the massless disk particles with different $\beta$ values in both our 250pl and 40pl models. In both cases, the larger disk particles with smaller $\beta$ values are concentrated closer to the star, while the smaller particles with larger $\beta$ values are more spread out as expected. For the 250pl model, which has an inner and outer ring component, the largest disk particles are most concentrated in the inner ring, while significantly less particles occupy the outer ring. As $\beta$ increases, the majority of disk particles are still concentrated in the inner ring, although a significant number of these small disk particles also populate the outer ring as well. 

\subsection{Radial Distribution vs. Planet Mass}
\begin{figure*}
    \centering
    \caption{\label{fig:1_0.5_mass} Radial density distribution of massless disk particles within 20 au of the disk midplane for our single 1 Jupiter mass model \textbf{left} and our single 0.5 Jupiter mass model \textbf{right}. The white dashed box highlights the location where the radial distribution of the disk particles differs between the two models. In both frames the density is scaled the same in log space.}
    \includegraphics[width=\textwidth]{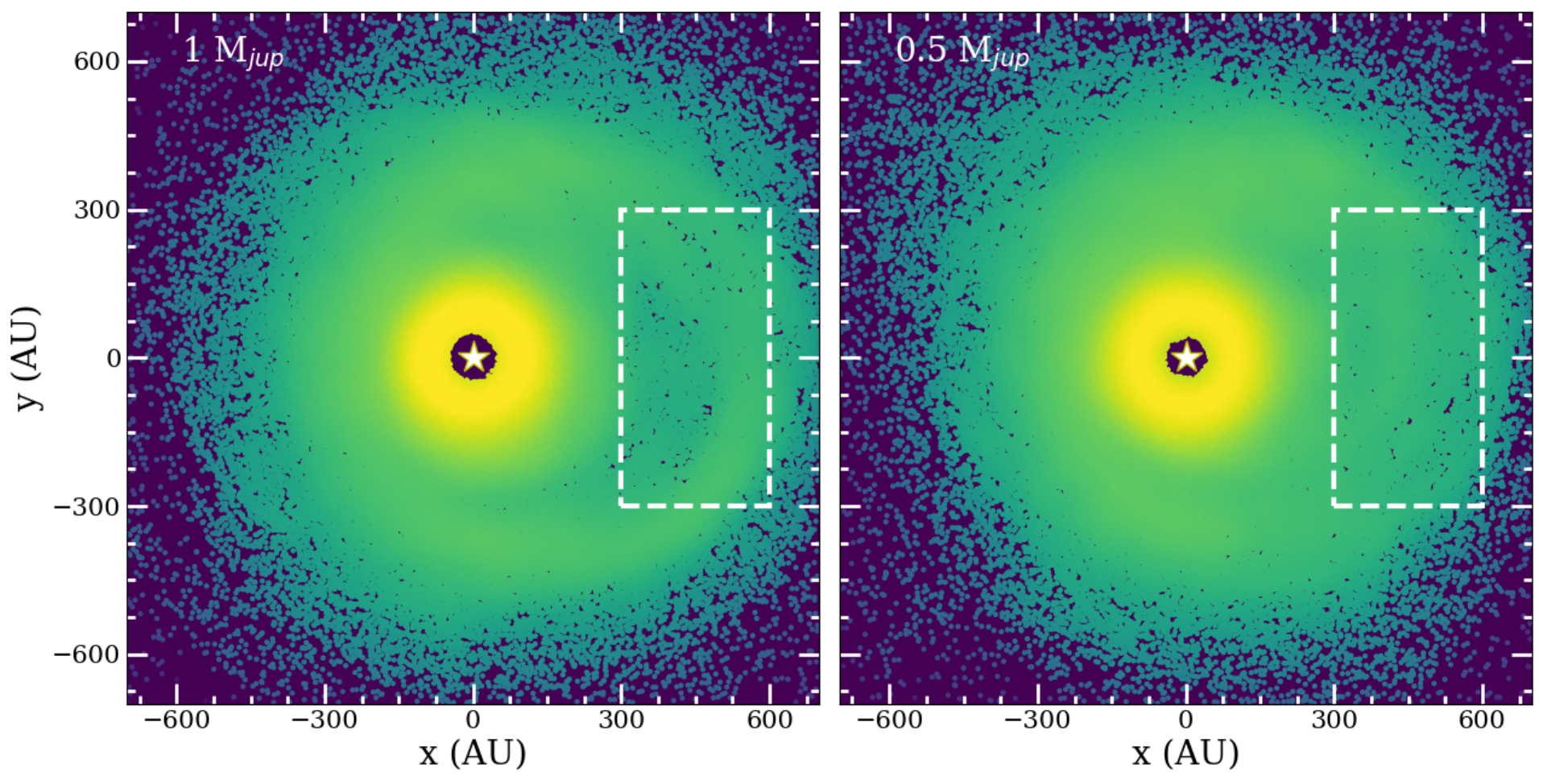}
\end{figure*}

in Figure \ref{fig:1_0.5_mass}, we show the radial distribution of disk particles within 20 au of the disk midplane for the 1 and 0.5 Jupiter planet mass models. For the 1 Jupiter mass model, we find that the gap is relatively more cleared of disk particles, where more particles are pushed out into the outer ring. In contrast, we find for the 0.5 Jupiter mass model that less disk particles are cleared from the gap, and consequently, less disk particles occupy the outer ring.

\clearpage

\bibliography{sample631}{}
\bibliographystyle{aasjournal}



\end{document}